\documentclass[aps]{revtex4}
\usepackage{amsfonts}
\usepackage{amsmath}
\usepackage{amssymb,epsf}

\begin{document}

\title{Extended phase space of Black Holes \\
in Lovelock gravity with nonlinear electrodynamics}
\author{S. H. Hendi$^{1,2}$\footnote{email address: hendi@shirazu.ac.ir}, S. Panahiyan$^{1}$\footnote{
email address: ziexify@gmail.com} and B. Eslam
Panah$^{1}$\footnote{email address:
behzad$_{-}$eslampanah@yahoo.com}} \affiliation{$^1$ Physics
Department and Biruni Observatory, College of Sciences, Shiraz
University, Shiraz 71454, Iran\\
$^{2}$ Research Institute for Astronomy and Astrophysics of Maragha (RIAAM),
Maragha, Iran}

\begin{abstract}
In this paper, we consider Lovelock gravity in presence of two
Born-Infeld types of nonlinear electrodynamics and study their
thermodynamical behavior. We extend the phase space by considering
cosmological constant as a thermodynamical pressure. We obtain
critical values of pressure, volume and temperature and
investigate the effects of both the Lovelock gravity and the
nonlinear electrodynamics on these values. We plot $P-v$, $T-v$
and $G-T$ diagrams to study the phase transition of these
thermodynamical systems. We show that power of the nonlinearity
and gravity have opposite effects. We also show how considering
cosmological constant, nonlinearity and Lovelock parameters as
thermodynamical variables will modify Smarr formula and first law
of thermodynamics. In addition, we study the behavior of universal
ratio of $\frac{P_{c}v_{c}}{T_{c}}$ for different values of
nonlinearity power of electrodynamics as well as the Lovelock
coefficients.
\end{abstract}

\maketitle

\section{introduction}

In context of AdS/CFT correspondence, it was proposed that
variation of cosmological constant corresponds to varying the
number of colors in the boundary field theory of Yang-Mills with
chemical potential interpretation \cite{Kast,John,Dolan1,Dolan2}.
On the other hand, according to Teitelboim and Henneaux's
mechanism, coupling four-dimensional gravity with antisymmetric
gauge field without cosmological constant results into
appearance of the cosmological constant as a constant of motion \cite%
{Teit}, and therefore, the cosmological constant will be a
variable. Recently, the cosmological constant was considered as a
state-dependant parameter in two dimensional dilaton gravity
\cite{Grumiller}. It was shown that treating the cosmological
constant as a $U(1)$ charge with non-minimal coupling leads to
confining electrostatic potential. Larranaga showed that
considering the cosmological constant as thermodynamical variable
can be extended to use the Smarr formula for inner and outer
horizons of BTZ black hole \cite{Larr}.

Recently, there has been an increasing interest in thermodynamical
behavior of black holes in asymptotically adS spacetime. This
growing interest comes from the fact that using adS/CFT
correspondence, one can find answers regarding a conformal field
theory of $d$-dimensions by solving problems that a gravitational
field presents in ($d+1$)-dimensional anti de Sitter spacetime
\cite{AdS/CFT}. Hawking and Page, in their pioneering work, showed
that the similarity of phase transition between the stable large
black hole and thermal gas in adS space can be interpreted as the
confinement/deconfinement phase transition in the dual strongly
coupled gauge theory \cite{HP}. Later, in an interesting article,
Witten showed that using adS/CFT correspondence, one can study the
thermal phase transition and the interpretation of confinement in
gauge theories \cite{Witten}. On the other hand, one may consider
the cosmological constant as a thermodynamical pressure (in order
to investigate the phase transition of black holes) to extend the
phase space (see \cite{KubizJHEP} for more details) and modify the
first law of black holes thermodynamics \cite{Dolan,Creighton}.
Another contribution of this consideration is a renewed
interpretation for the mass of black holes which from internal
energy becomes Enthalpy \cite{Kast}. This interpretation indicates
that the mass of black holes plays more important role in
thermodynamical structure of black holes and contains more
information regarding phase structure of black holes
\cite{Cvetic}.

For a canonical ensemble with a fixed charge, it was found that
there exists a phase transition between small and large black
holes. This phase transition behaves very like the gas/liquid
phase transition in a Van der Waals system \cite{Chamblin}. On the
other hand, phase transition of small/large black holes in adS/CFT
correspondence may be interpreted as conductor/superconductor
regions of the condescend matter systems \cite{Pan}.

Considering the fact that Maxwell theory contains some fundamental
problems and nonlinear electromagnetic fields solve some of these
shortcomings \cite{Born}, one is motivated to study different
models of nonlinear electrodynamics (NED). One interesting class
of these models is Born-Infeld (BI) type which is acquired in the
low energy limit of heterotic string theory \cite{BI-ST}.
Therefore, one is motivated to study these theories (which in this
paper we have considered logarithmic \cite{Soleng} and exponential
forms \cite{HendiEXP}) and the nonlinearity effects of
electromagnetic field on critical values representing phase
transition of black holes.

On the other hand, Einstein gravity is not flawless and has some fundamental
problems \cite{Stelle}. Generalization of the Einstein gravity to higher
orders of Lovelock gravity is one way to solve some of these problems \cite%
{Lovelock}. Besides, the Lagrangian of Lovelock gravity is obtainable
through the use of the low energy effective action of string theory \cite%
{BI-ST}. One can take this fact into account that modification of
Einstein gravity may change the conserved quantities of black
holes and therefore it is inevitable to see that critical values
and phase transition may depend on the choice of gravity model.

In literature, there have been various studies regarding higher
orders of Lovelock gravity in presence of different NED
\cite{higher-order}. Also, the phase transition and stability
conditions of black holes in various gravity models have been
studied intensively \cite{phase-higher}. These investigations lead
to interesting consequences and phenomenologies \cite{Phena}. In
this paper, we consider higher orders of Lovelock gravity in
presence of two classes of NED and study their phase structure. We
investigate the effects of both nonlinearity of the electrodynamic
models and Lovelock parameters on the phase diagrams and the
critical values.

Considering the fact that we are treating black holes as
thermodynamical system and interpret the first law of black holes
mechanics as the first law of thermodynamics, we expect to see the
similar thermodynamical behavior for the black holes and usual
thermodynamical systems. Therefore, it is crucial to investigate
phase transition and critical values of black holes. Moreover,
Lovelock gravity is a generalization of the Einstein gravity and
it is a theory that solves some of the Einstein gravity problems.
Hence, this generalization also gives a correction to calculated
critical values of the Einstein gravity. The thermodynamical
behavior of this modification should be reasonable and consistent
with thermodynamical concepts. Furthermore, as it was mentioned
before, nonlinear electromagnetic fields are introduced to solve
shortcomings of Maxwell theory. So it is reasonable to investigate
nonlinear effects of NED on the phase transition of black holes.
Also, modifications of electrodynamics like gravitational parts
must have consistent thermodynamical behavior.

This paper is constructed in the following form. Next section is
devoted to introduction to higher orders of Lovelock gravity and
conserved quantities. Then, we extend the phase space by
considering cosmological constant as thermodynamical pressure and
study the Smarr formula of these black holes. After that, we
calculate critical values and plot related diagrams for different
cases. We give a detailed discussion regarding diagrams, their
physical interpretations, and the effects of both NED and
gravitational parameters. We finish our paper with closing
remarks.

\section{Solutions and thermodynamic quantities of Lovelock gravity}

Here, we restrict ourselves to BI types NED models that were
introduced by Soleng \cite{Soleng} and Hendi \cite{HendiEXP} with
the following Lagrangians
\begin{equation}
\mathcal{L(F)}=\left\{
\begin{array}{c}
\beta ^{2}\left( \exp (-\frac{\mathcal{F}}{\beta ^{2}})-1\right) ,\;\;\text{%
ENEF} \\
-8\beta ^{2}\ln \left( 1+\frac{\mathcal{F}}{8\beta ^{2}}\right) ,\;\;\text{%
LNEF}%
\end{array}%
\right. , \label{LNED}
\end{equation}%
where ENEF and LNEF denote \emph{Exponential form of Nonlinear
Electromagnetic Field} and \emph{Logarithmic form of Nonlinear
Electromagnetic Field}, respectively, and $\beta $ is nonlinearity
parameter. Considering the fact that we are interested in studying
the spherically symmetric spacetime, we employ the following
metric
\begin{equation}
ds^{2}=-g(r)dt^{2}+\frac{dr^{2}}{g(r)}+r^{2}d\Omega _{d-2}^{2},
\label{Metric}
\end{equation}
where $d\Omega _{d}^{2}$ denotes the standard metric of
$d$-dimensional sphere, $S^{d}$, with the volume $\omega _{d}$. It
is known that the field equations of Einstein gravity in the
presence of NED are in the following forms \cite{HendiAnn}
\begin{eqnarray}
&&G_{\mu \nu }+\Lambda g_{\mu \nu }=\frac{1}{2}g_{\mu \nu
}\mathcal{L(F)}-2\frac{d\mathcal{L(F)}}{d\mathcal{F}}F_{\mu
\lambda }F_{\nu }^{\lambda }, \\ \label{EinsteinEq}
&&\nabla _{\mu
}\left( \frac{d\mathcal{L(F)}}{d\mathcal{F}}F^{\mu \nu }\right)
=0,  \label{MaxwellEq}
\end{eqnarray}
where $G_{\mu \nu }$ and $\Lambda$ are Einstein tensor and
cosmological constant, respectively. Einstein gravity in the
presence of mentioned NED has been studied in Ref.
\cite{HendiAnn}. Regardless of gravitational sector, one may
consider Eqs. (\ref{LNED}) and (\ref{MaxwellEq}) with the metric
(\ref{Metric}) to obtain the nonzero components of electromagnetic
fields with the following explicit forms \cite{HendiAnn}
\begin{equation}
E(r)=F_{tr}=\frac{q}{r^{2}}\times \left\{
\begin{array}{ll}
e^{-\frac{L_{W}}{2}}, & \text{ENEF} \\
\frac{2}{\Gamma +1}, & \text{LNEF}%
\end{array}
\right. ,  \label{E}
\end{equation}
where $q$ is an integration constant which is proportional to
total electric charge of the black hole solutions (with regarding
$4 \pi$ as proportionality constant) Extended phase space
thermodynamics and phase diagrams of the Einstein gravity in the
presence of NED were investigated in Ref. \cite{HPE}. So, we
consider Gauss-Bonnet (GB) and third order of Lovelock (TOL)
gravities and investigate extended phase space thermodynamics and
critical behavior of these gravities.

\subsection{GB gravity}

At first, we take into account the GB gravity. In order to obtain
the field equation of GB gravity, one should add the $G_{\mu \nu
}^{GB}$ tensor to the left hand side of Eq. (\ref{EinsteinEq}), in
which $G_{\mu \nu }^{GB}$ is
\begin{equation}
G_{\mu \nu }^{GB}=-\alpha _{GB}\left[ 4R^{\rho \sigma }R_{\mu \rho \nu
\sigma }-2R_{\mu }^{\rho \sigma \lambda }R_{\nu \rho \sigma \lambda
}-2RR_{\mu \nu }+4R_{\mu \lambda }R_{\nu }^{\lambda }+\frac{1}{2}g_{\mu \nu }%
\mathcal{L}_{GB}\right] ,  \label{GBterm}
\end{equation}%
where $\mathcal{L}_{GB}=R_{\mu \nu \gamma \delta }R^{\mu \nu
\gamma \delta }-4R_{\mu \nu }R^{\mu \nu }+R^{2}$ and $\alpha
_{GB}$ is GB parameter. Considering Eq. (\ref{EinsteinEq}) with
the extra term (\ref{GBterm}), one can obtain the following
solutions \cite{HendiP}
\begin{equation}
g_{GB}(r)=1+\frac{r^{2}}{2\alpha }\left( 1-\sqrt{\Psi_{GB} \left( r\right) }%
\right) ,  \label{gGB}
\end{equation}%
where $\alpha =\left( d-3\right) \left( d-4\right) \alpha _{GB}$ and%
\begin{equation}
\Psi_{GB} \left( r\right) =1+\frac{8\alpha \Lambda }{\left( d-1\right)
\left( d-2\right) }+\frac{4\alpha m}{r^{d-1}}+\frac{4\alpha \beta
^{2}\Upsilon }{\left( d-1\right) \left( d-2\right) },
\end{equation}
\begin{equation}
\Upsilon =\left\{
\begin{array}{cc}
1+\frac{2(d-1)q}{\beta r^{d-1}}\left[ \int \sqrt{L_{W}}dr-\int \frac{1}{%
\sqrt{L_{W}}}dr\right] ,&\;\;\text{ENEF} \\
\frac{8(d-2)}{\left( d-1\right) }\left[ \frac{\left( 2d-3\right) (\Gamma -1)%
}{d-2}-\frac{\left( d-1\right) \ln \left( \frac{1+\Gamma }{2}\right) }{d-2}+%
\frac{\left( d-2\right) \left( 1-\Gamma ^{2}\right)
\mathcal{H}}{d-3}\right] ,&\;\;\text{LNEF}
\end{array}%
\right.
\end{equation}%
in which $L_{W}=LambertW(\frac{4q^{2}}{\beta ^{2}r^{2d-4}})$ and
$m$ is an integration constant which is related to the total mass
\cite{HendiP}
\begin{equation}
M=\frac{\omega _{d-2}\left( d-2\right) m}{16\pi }.  \label{Mass}
\end{equation}%

In addition, $\mathcal{H}$ and $\Gamma $ are in the following
forms
\begin{eqnarray}
\mathcal{H} &=&_{2}{F}_{1}\left( \left[ \frac{1}{2},\frac{d-3}{2d-4}\right] ,%
\left[ \frac{3d-7}{2d-4}\right], 1-\Gamma ^{2} \right) , \nonumber \\
\Gamma &=&\sqrt{1+\frac{q^{2}}{\beta ^{2}r^{2d-4}}}. \nonumber
\end{eqnarray}%

Calculating the Kretschmann scalar, one finds that it diverges at
$r=0$, so the metric function (\ref{gGB}) has an essential
singularity at $r=0$ \cite{HendiP}. We should note that these
solutions may be interpreted as asymptotically adS black holes as
those of Einstein case \cite{HendiAnn}. Now, we take into account
the surface gravity interpretation to obtain the Hawking
temperature of the mentioned black hole solutions, yielding
\cite{HendiP}
\begin{equation}
T=\frac{-2\Lambda r_{+}^{4}+\left( d-2\right) \left( d-3\right)
r_{+}^{2}+\left( d-2\right) \left( d-5\right) \alpha -\varpi r_{+}}{4\pi
r_{+}\left( d-2\right) \left( r_{+}^{2}+2\alpha \right) },  \label{TempGB}
\end{equation}%
where%
\begin{equation}
\varpi =\left\{
\begin{array}{c}
\beta ^{2}r_{+}^{3}\left( \left[ 1+\left( \frac{2E}{\beta }\right) ^{2}%
\right] e^{\frac{-2E^{2}}{\beta ^{2}}}-1\right) ,\;\;\text{ENEF} \\
\;8r^{3}\beta ^{2}\ln \left[ 1-\left( \frac{E}{2\beta }\right) ^{2}\right] +%
\frac{4r^{3}E^{2}}{1-\left( \frac{E}{2\beta }\right) ^{2}},\;\;\text{LNEF}%
\end{array}%
\right. ,
\end{equation}%
and $E=\left. E(r)\right\vert _{r=r_{+}}$. Since obtained solutions are
asymptotically adS, one may obtain the entropy of the black hole solutions
by the use of the Gibbs-Duhem relation. After some calculations, one can
obtain \cite{HendiP}
\begin{equation}
S=\frac{\omega_{d-2}}{4}\left( r_{+}^{d-2}+\frac{2\left(
d-2\right) }{\left( d-4\right) }\alpha r_{+}^{d-4}\right) ,
\label{EntropyGB}
\end{equation}%
which confirms that the area law is recovered for $\alpha =0$.

\subsection{TOL gravity}

Now, we insert the following TOL term, $G_{\mu \nu }^{TOL}$, to the field
equation of GB gravity to obtain the solutions of TOL gravity. The tensor $%
G_{\mu \nu }^{TOL}$ may be written as
\begin{eqnarray}
G_{\mu \nu }^{TOL} &=&-\alpha _{TOL}[3(4R^{\tau \rho \sigma \kappa
}R_{\sigma \kappa \lambda \rho }R_{\nu \tau \mu }^{\lambda }-8R_{\lambda
\sigma }^{\tau \rho }R_{\tau \mu }^{\sigma \kappa }R_{\nu \rho \kappa
}^{\lambda }+2R_{\nu }^{\tau \sigma \kappa }R_{\sigma \kappa \lambda \rho
}R_{\tau \mu }^{\lambda \rho }-R^{\tau \rho \sigma \kappa }R_{\sigma \kappa
\tau \rho }R_{\nu \mu }  \nonumber \\
&&+8R_{\nu \sigma \rho }^{\tau }R_{\tau \mu }^{\sigma \kappa
}R_{\kappa }^{\rho }+8R_{\nu \tau \kappa }^{\sigma }R_{\sigma \mu
}^{\tau \rho }R_{\rho }^{\kappa }+4R_{\nu }^{\tau \sigma \kappa
}R_{\sigma \kappa \mu \rho }R_{\tau }^{\rho }-4R_{\nu }^{\tau
\sigma \kappa }R_{\sigma \kappa \tau \rho }R_{\mu
}^{\rho}+4R^{\tau \rho \sigma \kappa }R_{\sigma \kappa \tau \mu
}R_{\nu \rho}  \nonumber \\
&&+2RR_{\nu }^{\kappa \tau \rho }R_{\tau \rho \kappa \mu }+8R_{\nu
\mu \rho }^{\tau }R_{\sigma }^{\rho }R_{\tau }^{\sigma }-8R_{\nu
\tau \rho }^{\sigma }R_{\sigma }^{\tau }R_{\mu }^{\rho}-8R_{\sigma
\mu }^{\tau \rho }R_{\tau }^{\sigma }R_{\nu \rho
}-4RR_{\nu \mu \rho }^{\tau }R_{\tau }^{\rho }  \nonumber \\
&&+4R^{\tau \rho }R_{\rho \tau }R_{\nu \mu }-8R_{\nu }^{\tau
}R_{\tau \rho }R_{\mu }^{\rho }+4RR_{\nu \rho }R_{\mu }^{\rho
}-R^{2}R_{\nu \mu })+\frac{1}{2}g_{\mu \nu }\mathcal{L}_{TOL}],
\label{TOLterm}
\end{eqnarray}%
where $\alpha _{TOL}$ and $\mathcal{L}_{TOL}$ are, respectively, the
coefficient and Lagrangian of TOL gravity
\begin{eqnarray}
\mathcal{L}_{TOL} &=& 2R^{\lambda \varepsilon \sigma \kappa}
R_{\sigma \kappa \rho \tau} R^{\rho \tau}_{\lambda \varepsilon}
+8R_{\sigma \rho }^{\mu \nu }R_{\nu \tau }^{\sigma \kappa }R_{\mu
\kappa }^{\rho \tau }+24R^{\mu \nu \sigma \kappa }R_{\sigma \kappa
\nu \rho }R_{\mu }^{\rho }+3RR^{\mu \nu \sigma \kappa
}R_{\sigma \kappa \mu \nu }  \nonumber \\
&&-12RR_{\mu \nu }R^{\mu \nu }+24R^{\mu \nu \sigma \kappa }R_{\sigma \mu
}R_{\kappa \nu }+16R^{\mu \nu }R_{\nu \sigma }R_{\mu }^{\sigma }+R^{3}.
\label{LTOL}
\end{eqnarray}

Hereafter, we consider the special case $\alpha _{TOL}=\frac{\left(
d-3\right) \left( d-4\right) }{3\left( d-5\right) \left( d-6\right) }\alpha
_{GB}^{2}$ to simplify the calculations. It has been shown that the metric
function of TOL gravity in the presence of NED can be written as \cite%
{HendiDehghani}%
\begin{equation}
g_{_{TOL}}(r)=1+\frac{r^{2}}{\alpha }\left( 1- \Psi_{TOL} \left( r\right) ^{%
\frac{1}{3}}\right) ,  \label{gTOL}
\end{equation}

\begin{equation}
\Psi_{TOL} \left( r\right)=1+\frac{6\alpha \Lambda }{\left( d-1\right)
\left( d-2\right) }+\frac{3\alpha m}{r^{d-1}}+\frac{3\alpha \beta
^{2}\Upsilon }{\left( d-1\right) \left( d-2\right) }.
\end{equation}

The geometric and thermodynamic properties of the asymptotically
adS black holes have been studied before \cite{HendiDehghani}. The
finite mass of these solutions is the same as that of Einstein
gravity, where $m$ can be obtained as a function of $r_{+}$ from
the metric function of TOL gravity. The Hawking temperature and
the entropy of the TOL solutions can be calculated as
\cite{HendiDehghani}
\begin{equation}
T=\frac{-6\Lambda r_{+}^{6}+3\left( d-2\right) \left( d-3\right)
r_{+}^{4}+3\left( d-2\right) \left( d-5\right) \alpha r_{+}^{2}+\left(
d-2\right) \left( d-7\right) \alpha ^{2}-3\varpi r_{+}^{3}}{12\pi
r_{+}\left( d-2\right) \left( r_{+}^{2}+\alpha \right) ^{2}},
\end{equation}%
and
\begin{equation}
S=\frac{\omega_{d-2}}{4}\left( r_{+}^{d-2}+\frac{2\left(
d-2\right) }{\left(
d-4\right) }\alpha r_{+}^{d-4}+\frac{\left( d-2\right) }{\left( d-6\right) }%
\alpha ^{2}r_{+}^{d-6}\right) .
\end{equation}

\section{Extended phase space and phase diagrams}

As we mentioned in introduction, there are some motivations to
view the cosmological constant as a variable. In addition, there
are various theories in which some physical constants (such as
gauge coupling constants, Newton constant, Lovelock coefficients
and BI parameter) are not fixed but dynamical ones. In that case,
it is logical to consider the variation of
these parameters into the first law of black hole thermodynamics \cite%
{CteVariable}.

In order to investigate the phase structure of these classes of
gravities, we employ the approach in which the cosmological
constant is a thermodynamical variable (pressure) with the
following relation
\begin{equation}
P=-\frac{\Lambda }{8\pi }. \label{P}
\end{equation}

This consideration could be justified due to the fact that in
quantum context, fundamental fixed parameters could vary. As one
can see the conjugating thermodynamical variable to this
assumption (cosmological constant as thermodynamical pressure)
will be volume where in literature the derived volume for
different types of black holes are in agreement with the topology
of the spacetime \cite{Cvetic,phase-higher,HPE}. In order to
calculate the volume of these thermodynamical systems we use the
following relation
\begin{equation}
V=\left( \frac{\partial H}{\partial P}\right) _{S,Q}.
\end{equation}

Also, we should consider the effects of cosmological constant in the first
law of thermodynamics and extend our phase space. With doing so the total
finite mass of the black hole will play the role of Enthalpy and hence the
corresponding Gibbs free energy will be in form of
\begin{equation}
G=H-TS=M-TS. \label{G}
\end{equation}

Using the mentioned comments, one can obtain the volume with the
following form
\begin{equation}
V=\frac{\omega _{d-2}r_{+}^{d-1}}{d-1},
\end{equation}%
which is consistent with topological structure of spherical
symmetric spacetime. Equation (\ref{G}) was obtained in Einstein
gravity \cite{HPE} which indicates that although considering
Lovelock gravity modifies the metric function and some conserved
quantities of the black hole, it does not change the volume of the
black hole.

In addition, it was shown that the Smarr formula may be extended
to Lovelock gravity as well as nonlinear theories of
electrodynamics \cite{SmarrNew,CaiGB}. Geometrical techniques
(scaling argument) were used to derive an extension of the first
law and its related modified Smarr formula. The result includes
variations in the cosmological constant, Lovelock coefficients and
also nonlinearity parameter. In our case, Lovelock gravity in the
presence of NED, $M$ should be the function of entropy, pressure,
charge, Lovelock parameters and BI coupling coefficient
\cite{SmarrNew}. Regarding the previous section, we find that
those thermodynamic quantities satisfy the following differential
form
\begin{equation}
dM=TdS+\Phi dQ+VdP+\mathcal{A}_{1}^{\prime }d\alpha _{2}+\mathcal{A}%
_{2}^{\prime }d\alpha _{3}+\mathcal{B}d\beta ,  \label{GenFirstLaw}
\end{equation}
where we have achieved $T$, and one can obtain
\begin{eqnarray*}
\Phi&=& \left( \frac{\partial M}{\partial Q}\right) _{S,P,\alpha
_{2},\alpha
_{3},\beta },\\
V &=&\left( \frac{\partial M}{\partial P}\right) _{S,Q,\alpha
_{2},\alpha
_{3},\beta }, \\
\mathcal{A}_{1}^{\prime } &=&\left( \frac{\partial M}{\partial \alpha _{2}}%
\right) _{S,Q,P,\alpha _{3},\beta }, \\
\mathcal{A}_{2}^{\prime } &=&\left( \frac{\partial M}{\partial \alpha _{3}}%
\right) _{S,Q,P,\alpha _{2},\beta }, \\
\mathcal{B} &=&\left( \frac{\partial M}{\partial \beta }\right)
_{S,Q,P,\alpha _{2},\alpha _{3}}.
\end{eqnarray*}

Using the redefinition of $\alpha _{2}$ and $\alpha _{3}$ with respect to
the single parameter, $\alpha $, we can rewrite $\mathcal{A}_{1}^{\prime
}d\alpha _{2}+\mathcal{A}_{2}^{\prime }d\alpha _{3}$ as a single
differential form
\begin{eqnarray*}
d{\alpha _{2}} &=&\frac{1}{{(d-3)(d-4)}}d\alpha , \\
d\alpha _{3} &=&\frac{2{\alpha }}{{3(d-3)(d-4)(d-5)(d-6)}}{d\alpha .}
\end{eqnarray*}

Moreover, by scaling argument, we can obtain the generalized Smarr relation
for our asymptotically adS solutions in the extended phase space
\begin{equation}
(d-3)M=(d-2)TS+(d-3)Q\Phi -2PV+2\left( \mathcal{A}_{1}\alpha +\mathcal{A}%
_{2}\alpha ^{2}\right) -\mathcal{B}\beta,   \label{Smarr2}
\end{equation}%
where
\begin{eqnarray*}
\Phi &=&\left\{
\begin{array}{cc}
\frac{{\beta {r_{+}}\sqrt{L{w_{+}}}}}{{2(d-3)(3d-7)}}\left[
{(d-2)\,\mathcal{D}_{+}{L}_{W+}+3d-7}\right] , & \text{ENED} \\
-\frac{{2{\beta ^{2}}r_{+}^{d-1}}}{{(d-1)q}}\left( {{\eta
_{+}}-1}\right) , &
\text{LNED}%
\end{array}%
\right. ,
\end{eqnarray*}
\begin{eqnarray*}
V &=&\frac{r_{+}^{d-1}}{\left( d-1\right) }, \\
\mathcal{A}_{1} &=&\frac{(d-2)k^{2}r_{+}^{d-5}}{16\pi }-\frac{%
(d-2)kTr_{+}^{d-4}}{2(d-4)}, \\
\mathcal{A}_{2} &=&\frac{(d-2)k^{3}r_{+}^{d-7}}{24\pi }-\frac{%
(d-2)k^{2}Tr_{+}^{d-6}}{2(d-6)}, \\
\left. \mathcal{B}\right\vert _{\text{ENED}}
&=&\frac{q(d-2)r_{+}\left( L_{W+}\right)
^{\frac{3}{2}}\mathcal{D}_{+}}{8\pi \left(
d-1\right) (3d-7)}-\frac{\beta r_{+}^{d-1}}{8\pi \left( d-1\right) }+\frac{%
q\beta r_{+}^{d}\sqrt{L_{W+}}\left( 1-L_{W+}\right) }{8\pi \left( d-1\right)
\left( 1+L_{W+}\right) }+\frac{2qr_{+}}{8\pi \left( d-1\right) \sqrt{L_{W+}}%
\left( 1+L_{W+}\right) }, \\
\left. \mathcal{B}\right\vert _{\text{LNED}} &=&\frac{\beta r_{+}^{d-1}}{%
2\pi \left( d-1\right) ^{2}}\left[ (d-2)\left( \Gamma _{+}^{2}-1\right)
\mathcal{H}_{+}+2\left( d-1\right) \ln \left( \frac{1+\Gamma _{+}}{2}\right)
+(3d-5)\left( 1-\Gamma _{+}\right) \right] ,
\end{eqnarray*}%
in which

\begin{eqnarray*}
\eta_{+}&=&{}_{2}{F_{1}}\left( {\left[ {-\frac{1}{2},\,\frac{{1-d}}{{2d-4}}}%
\right] ,\,\left[ {\frac{{d-3}}{{2d-4}}}\right] ,\,1-}\Gamma
_{+}^{2}\right), \\
\mathcal{D}_{+}&=&_{1}{F}_{1} \left( [1],\left[\frac{5d-11}{2d-4}\right],\frac{%
L_{W+}}{2d-4}\right), \\
\mathcal{H}_{+}&=&_{2}{F}_{1}\left( \left[
\frac{1}{2},\frac{d-3}{2d-4}\right] ,\left[
\frac{3d-7}{2d-4}\right] , 1-\Gamma _{+}^{2} \right).
\end{eqnarray*}


Next step will be calculating critical values. Due to relation between
volume and radius of the black hole, we use horizon radius (specific volume)
in order to investigate the critical behavior of these systems \cite%
{phase-higher}. In order to do so, we use the method in which critical
values are obtained through the use of $P-r_{+}$ diagrams. At first, we use
the following relations to obtain the proper equations for critical radius
\begin{equation*}
\left( \frac{\partial P}{\partial r_{+}}\right) =\left( \frac{\partial ^{2}P%
}{\partial r_{+}^{2}}\right) =0.
\end{equation*}

For the economical reasons we will not bring obtained relations
for calculating critical horizon radius. We employ numerical
method for calculating critical values which result into following
diagrams for different classes of Lovelock gravity. We present
various tables in order to plot $P-r_{+}$, $T-r_{+}$ and $G-T$ and
study the effects of gravitational parameter which is presented by
$\alpha$ and NED parameter which is presented by $\beta$. In this
paper, we have considered two classes of NED. Studying different
phase diagrams for these NED shows that they have similar
behavior. Therefore, for economical reason, we will regard only
LNEF branch and calculate related critical values of this NED
model.

It will be constructive to give a short description regarding to different
phase diagrams and the information they contain before presenting tables and
phase diagrams. $G-T$ diagrams are representing energy level of different
states that phase transition takes place between them and shows the changes
in energy level of before and after phase transition states. The
characteristic swallowtail that is seen in these diagrams shows the process
that we know as phase transition. It also gives interesting information
regarding temperature of critical points. For $T-r_{+}$ plot, it contains
information regarding critical temperature and horizon radius in which phase
transition takes place. Also, it gives some insight about single state
regions which in our case is small/large black holes. Finally, studying $%
P-r_{+}$ plot gives us information regarding the behavior of
pressure as function of horizon radius, critical pressure and
critical horizon radius (volume) of phase transition. One of the
reasons for studying these diagrams is the similarity between
phase structure of black holes and the Van der Waals
thermodynamical systems.

In what follows, we present various tables to investigate the
effects of electrodynamics and gravity models on the critical
values of phase transition. We also plot $P-r_{+}$, $T-r_{+}$ and
$G-T$ diagrams for GB and TOL gravities and interpret them. It is
notable to mention that considering the metric function of GB
gravity, one finds that there is an upper limit for GB parameter
to have a real solution.


\begin{center}
\begin{tabular}{ccccc}
\hline\hline
$\beta $ & $v_{c}$ & $T_{c}$ & $P_{c}$ & $\frac{P_{c}v_{c}}{T_{c}}$ \\
\hline\hline
$0.10000$ & $1.03178$ & $0.16302$ & $0.03904$ & $0.24708$ \\ \hline
$0.50000$ & $1.60193$ & $0.13789$ & $0.02536$ & $0.29467$ \\ \hline
$1.00000$ & $1.63292$ & $0.13703$ & $0.02497$ & $0.29765$ \\ \hline
$1.50000$ & $1.63852$ & $0.13687$ & $0.02490$ & $0.29818$ \\ \hline
$2.00000$ & $1.64047$ & $0.13682$ & $0.02488$ & $0.29837$ \\ \hline
\end{tabular}
\\[0pt]
\vspace{0.1cm} Table ($1$): GB gravity with $q=1$, $\alpha =0.1$ and $d=5$.
\vspace{0.5cm}
\end{center}

\begin{center}
\begin{tabular}{ccccc}
\hline\hline
$\alpha $ & $v_{c}$ & $T_{c}$ & $P_{c}$ & $\frac{P_{c}v_{c}}{T_{c}}$ \\
\hline\hline
$0.10000$ & $1.63292$ & $0.13703$ & $0.02497$ & $0.29765$ \\ \hline
$0.30000$ & $1.89452$ & $0.10325$ & $0.01531$ & $0.28094$ \\ \hline
$0.50000$ & $2.12876$ & $0.08559$ & $0.01089$ & $0.27105$ \\ \hline
$0.70000$ & $2.34857$ & $0.07441$ & $0.00839$ & $0.26488$ \\ \hline
\end{tabular}
\\[0pt]
\vspace{0.1cm} Table ($2$): GB gravity with $q=1$, $\beta =1$ and $d=5$.
\vspace{0.5cm}
\end{center}

\begin{center}
\begin{tabular}{ccccc}
\hline\hline
$\beta $ & $v_{c}$ & $T_{c}$ & $P_{c}$ & $\frac{P_{c}v_{c}}{T_{c}}$ \\
\hline\hline
$0.10000$ & $0.91666$ & $0.40408$ & $0.20018$ & $0.45410$ \\ \hline
$0.50000$ & $1.24445$ & $0.36343$ & $0.15122$ & $0.51783$ \\ \hline
$1.00000$ & $1.26635$ & $0.36161$ & $0.14932$ & $0.52292$ \\ \hline
$1.50000$ & $1.27031$ & $0.36128$ & $0.14898$ & $0.52383$ \\ \hline
$2.00000$ & $1.27169$ & $0.36117$ & $0.14883$ & $0.52404$ \\ \hline
\end{tabular}
\\[0pt]
\vspace{0.1cm} Table ($3$): GB gravity with $q=1$, $\alpha =0.1$ and $d=7$.%
\vspace{0.5cm}
\end{center}

\begin{center}
\begin{tabular}{ccccc}
\hline\hline
$\alpha $ & $v_{c}$ & $T_{c}$ & $P_{c}$ & $\frac{P_{c}v_{c}}{T_{c}}$ \\
\hline\hline
$0.10000$ & $1.26635$ & $0.36161$ & $0.14932$ & $0.52292$ \\ \hline
$0.30000$ & $1.35716$ & $0.26569$ & $0.09268$ & $0.47341$ \\ \hline
$0.50000$ & $1.40740$ & $0.21949$ & $0.06844$ & $0.43886$ \\ \hline
$0.70000$ & $1.42508$ & $0.19154$ & $0.05528$ & $0.41136$ \\ \hline
\end{tabular}
\\[0pt]
\vspace{0.1cm} Table ($4$): GB gravity with $q=1$, $\beta =1$ and $d=7$.%
\vspace{0.5cm}
\end{center}

\begin{center}
\begin{tabular}{ccccc}
\hline\hline
$\beta $ & $v_{c}$ & $T_{c}$ & $P_{c}$ & $\frac{P_{c}v_{c}}{T_{c}}$ \\
\hline\hline
$0.10000$ & $0.97952$ & $0.38964$ & $0.18323$ & $0.46064$ \\ \hline
$0.50000$ & $1.25908$ & $0.35816$ & $0.14674$ & $0.51587$ \\ \hline
$1.00000$ & $1.27837$ & $0.35665$ & $0.14518$ & $0.52038$ \\ \hline
$1.50000$ & $1.28190$ & $0.35637$ & $0.14489$ & $0.52120$ \\ \hline
$2.00000$ & $1.28313$ & $0.35628$ & $0.14479$ & $0.52148$ \\ \hline
\end{tabular}
\\[0pt]
\vspace{0.1cm} Table ($5$): TOL gravity with $q=1$, $\alpha =0.1$ and $d=7$.%
\vspace{0.5cm}
\end{center}

\begin{center}
\begin{tabular}{ccccc}
\hline\hline
$\alpha $ & $v_{c}$ & $T_{c}$ & $P_{c}$ & $\frac{P_{c}v_{c}}{T_{c}}$ \\
\hline\hline
$0.10000$ & $1.27837$ & $0.35665$ & $0.14518$ & $0.52038$ \\ \hline
$0.30000$ & $1.45492$ & $0.25053$ & $0.08023$ & $0.46593$ \\ \hline
$0.50000$ & $1.66940$ & $0.19960$ & $0.05256$ & $0.43959$ \\ \hline
$0.70000$ & $1.90580$ & $0.16970$ & $0.03830$ & $0.43016$ \\ \hline
\end{tabular}
\\[0pt]
\vspace{0.1cm} Table ($6$): TOL gravity with $q=1$, $\beta =1$ and $d=7$.
\end{center}

\begin{figure}[tbp]
$%
\begin{array}{ccc}
\epsfxsize=5cm \epsffile{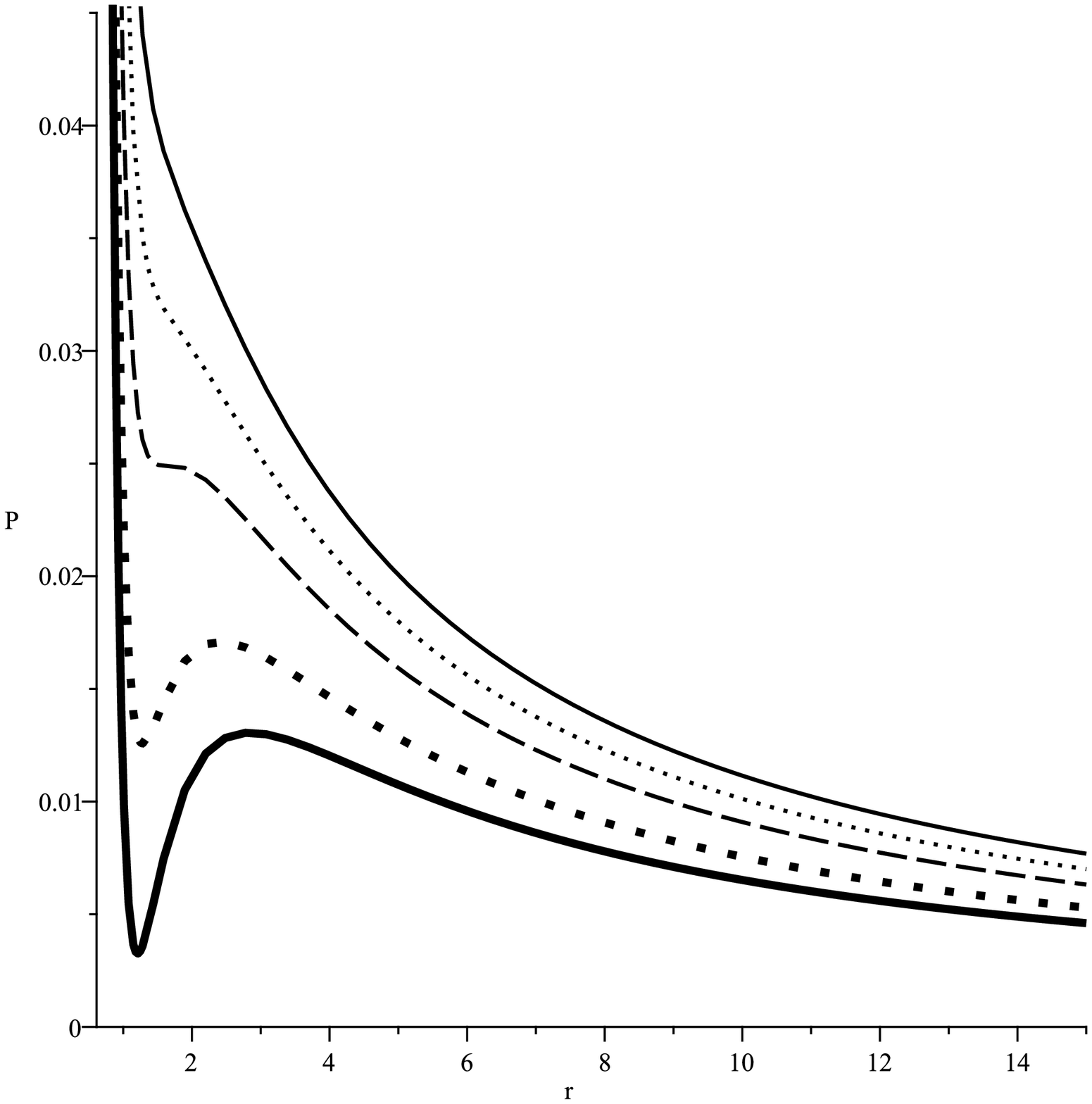} & \epsfxsize=5cm %
\epsffile{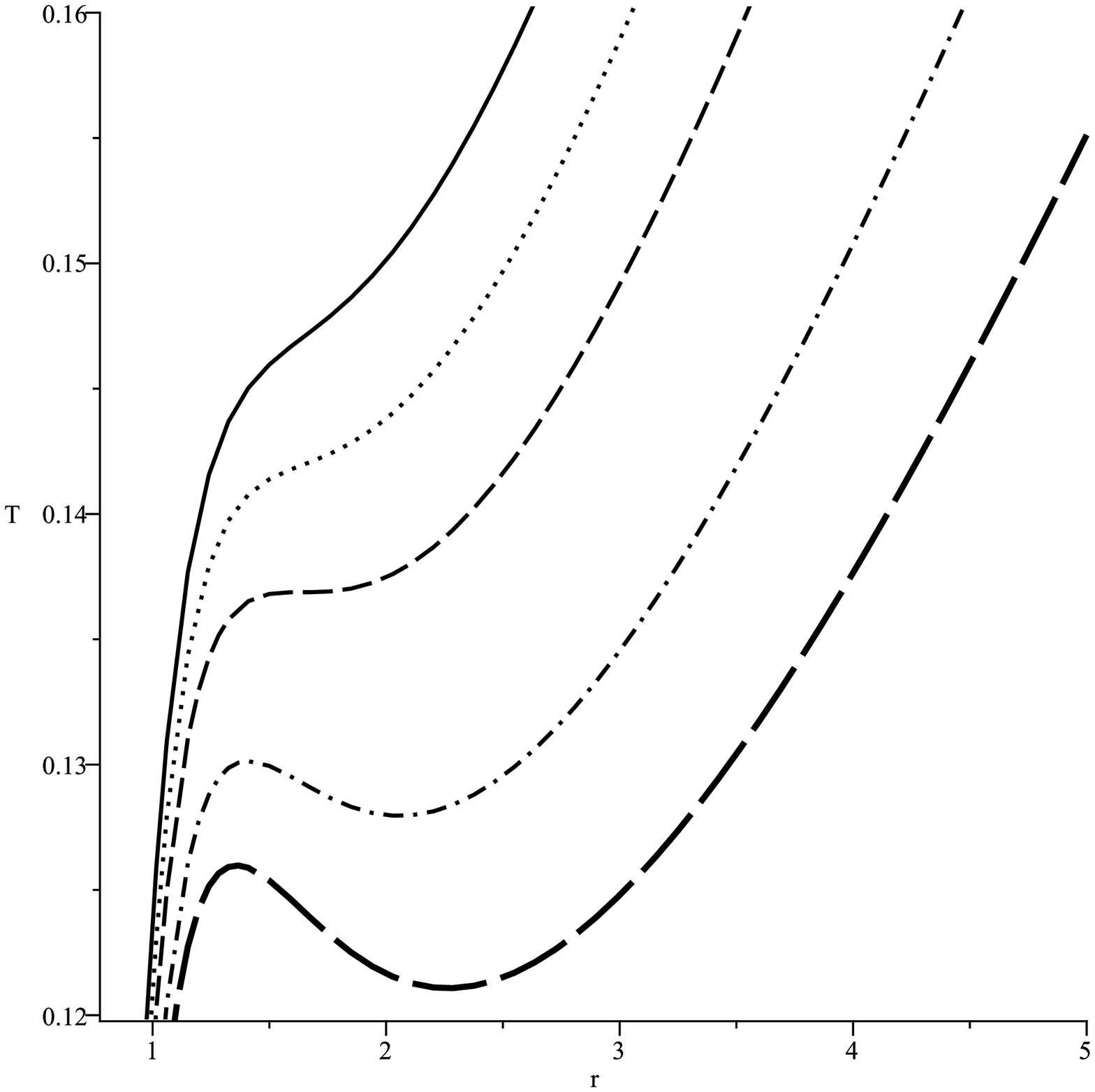} & \epsfxsize=5cm %
\epsffile{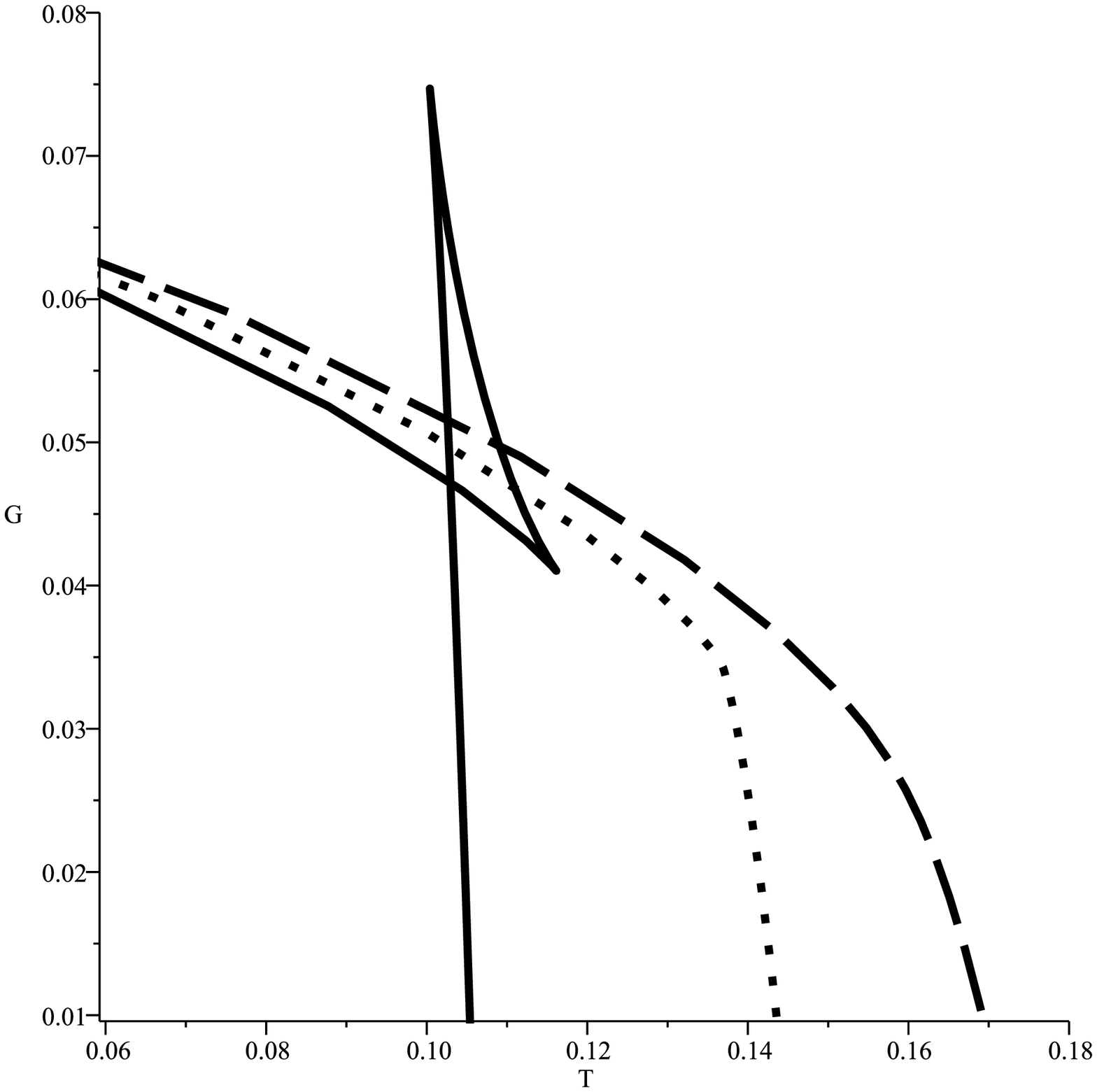}%
\end{array}
$%
\caption{ $P-v$ (left), $T-v$ (middle) and $G-T$ (right) diagrams in GB
gravity for $\protect\alpha=0.1$, $q=1$ and $\protect\beta=1.5$ ($d=5$).
\newline
$P-v$ diagram, from up to bottom $T=1.2T_{c}$, $T=1.1T_{c}$, $T=T_{c}$, $%
T=0.85T_{c}$ and $T=0.75T_{c}$, respectively. \newline
$T-v$ diagram, from up to bottom $P=1.2P_{c} $, $P=1.1P_{c}$. $P=P_{c}$, $%
P=0.85P_{c}$ and $P=0.75P_{c}$, respectively. \newline $G-T$
diagram for $P=0.5P_{c}$ (continuous line), $P=P_{c}$ (dotted
line) and $P=1.5P_{c}$ (dashed line). } \label{Fig1}
\end{figure}
\begin{figure}[tbp]
$%
\begin{array}{ccc}
\epsfxsize=5cm \epsffile{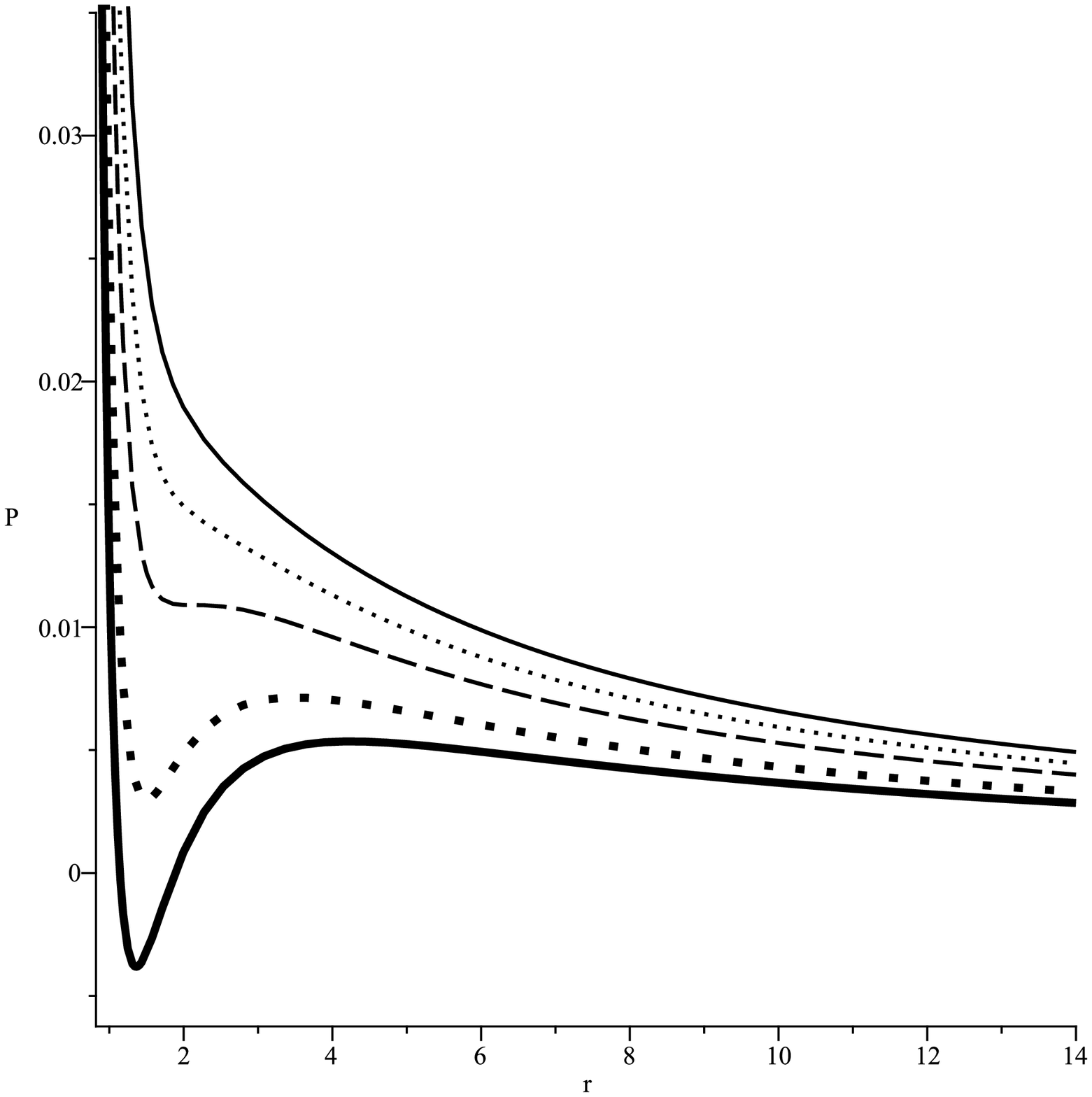} & \epsfxsize=5cm %
\epsffile{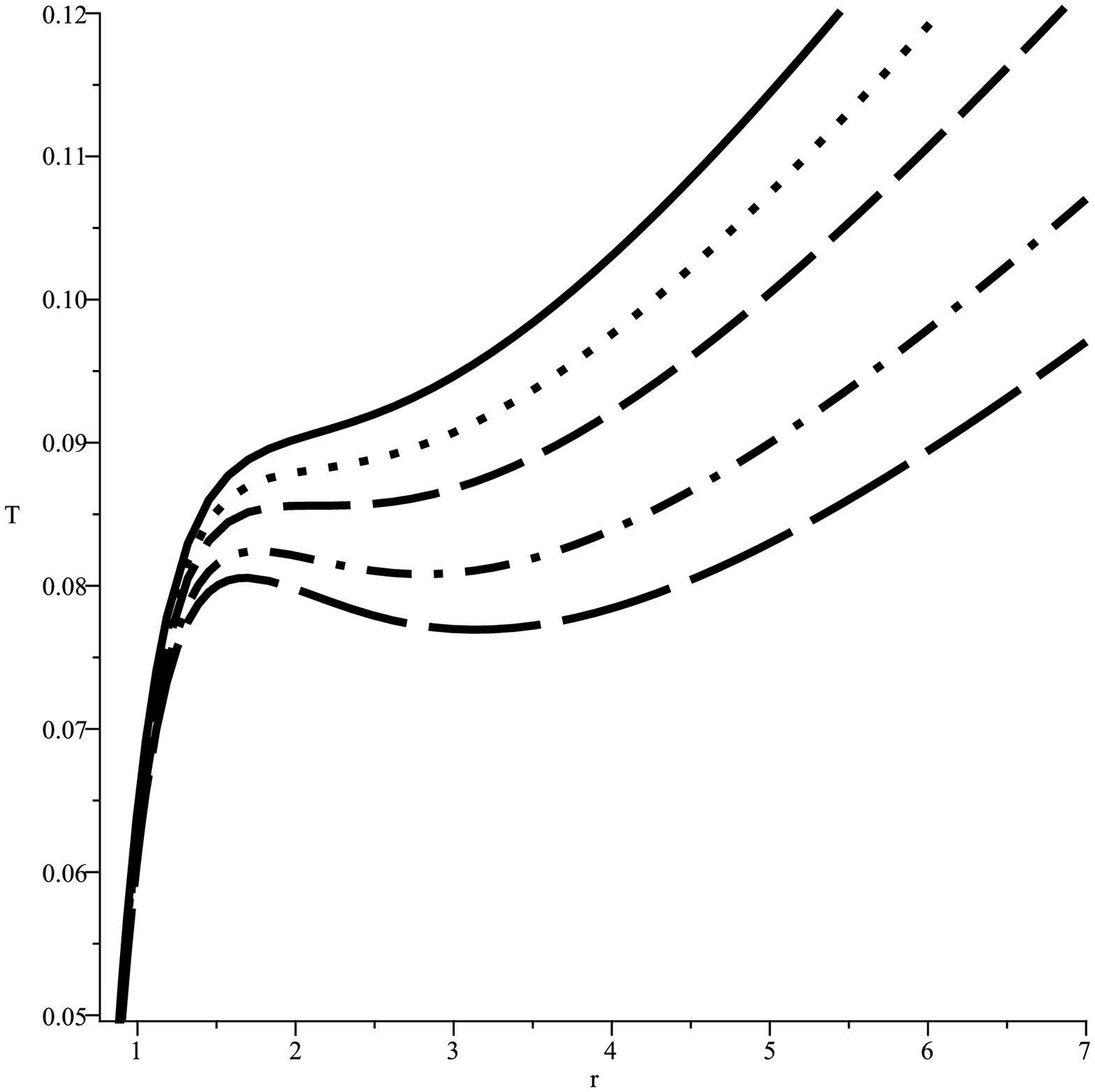} & \epsfxsize=5cm %
\epsffile{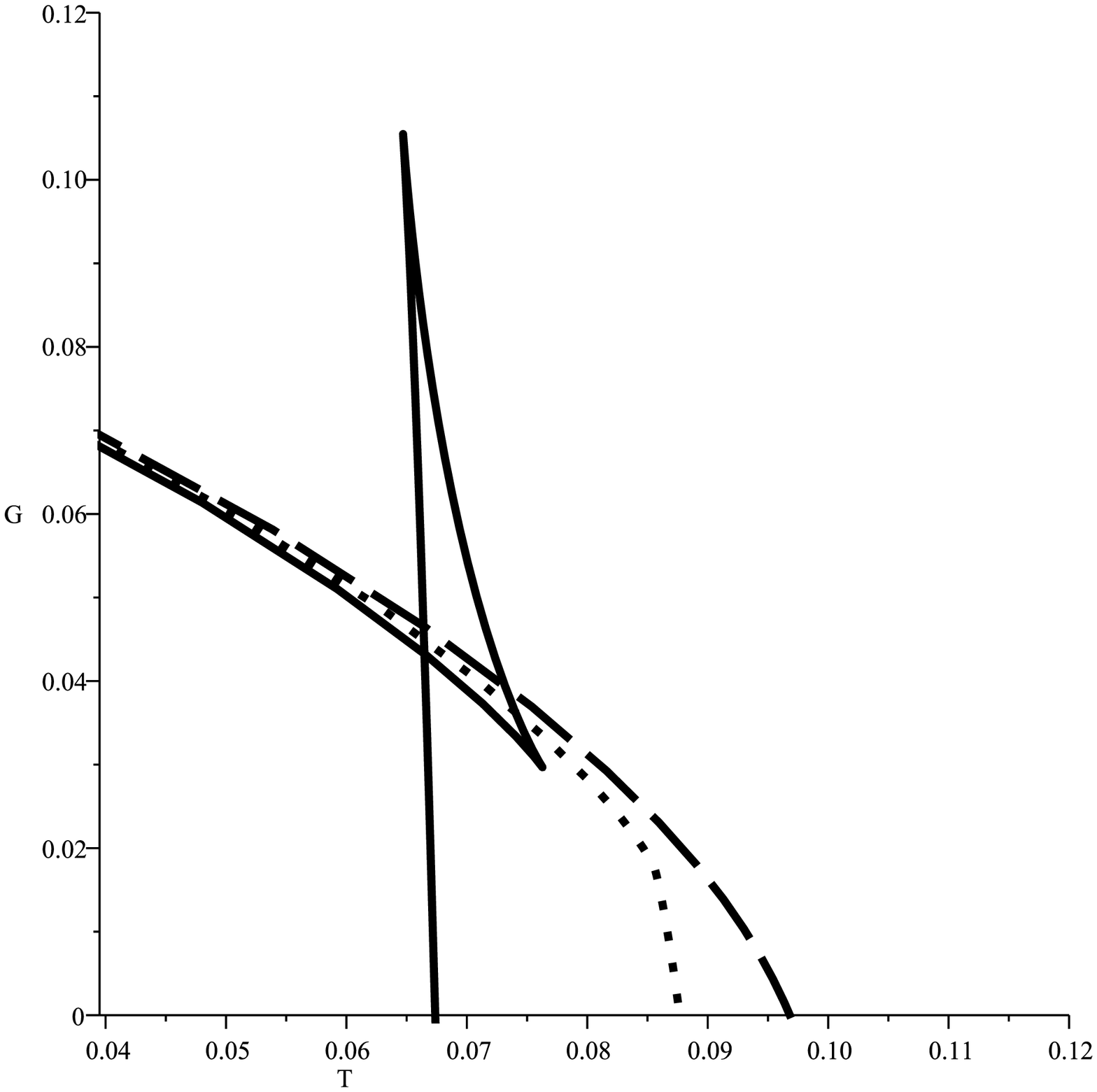}%
\end{array}
$%
\caption{ $P-v$ (left), $T-v$ (middle) and $G-T$ (right) diagrams in GB
gravity for $\protect\beta=1$, $q=1$ and $\protect\alpha=0.5$ ($d=5$).
\newline
$P-v$ diagram, from up to bottom $T=1.2T_{c}$, $T=1.1T_{c}$, $T=T_{c}$, $%
T=0.85T_{c}$ and $T=0.75T_{c}$, respectively. \newline
$T-v$ diagram, from up to bottom $P=1.2P_{c} $, $P=1.1P_{c}$. $P=P_{c}$, $%
P=0.85P_{c}$ and $P=0.75P_{c}$, respectively. \newline $G-T$
diagram for $P=0.5P_{c}$ (continuous line), $P=P_{c}$ (dotted
line) and $P=1.5P_{c}$ (dashed line). } \label{Fig2}
\end{figure}

\begin{figure}[tbp]
$%
\begin{array}{ccc}
\epsfxsize=5cm \epsffile{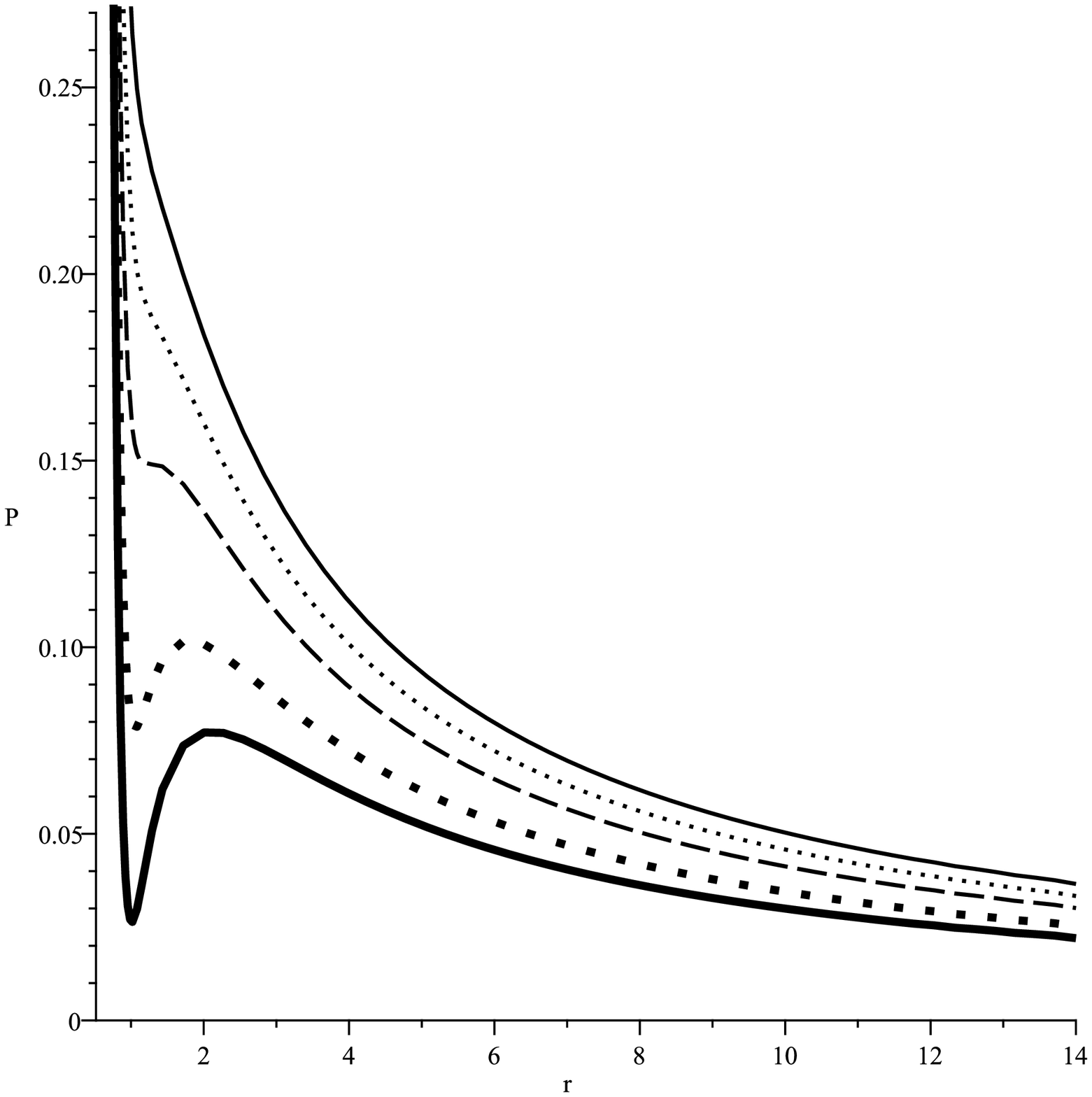} & \epsfxsize=5cm %
\epsffile{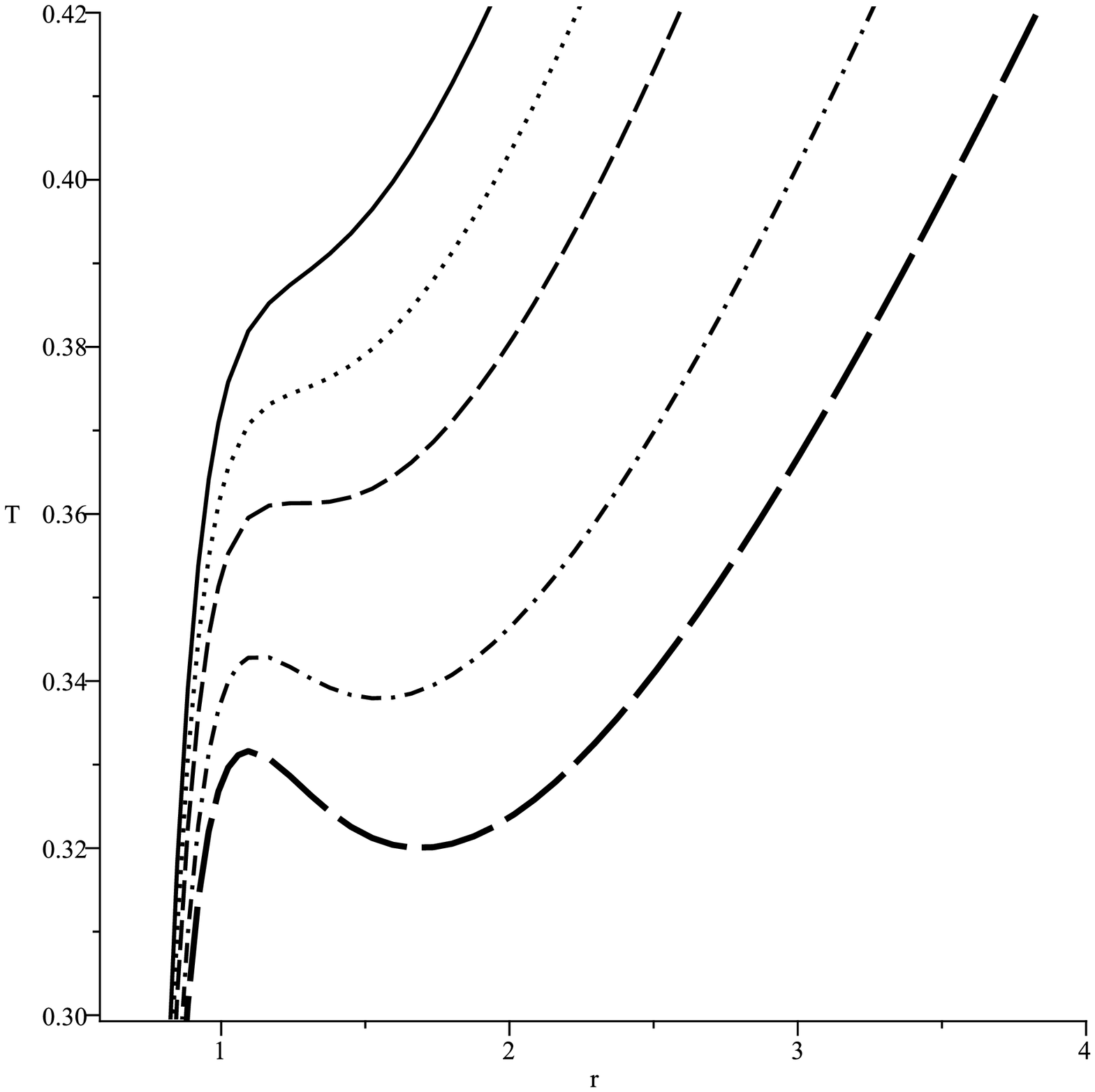} & \epsfxsize=5cm %
\epsffile{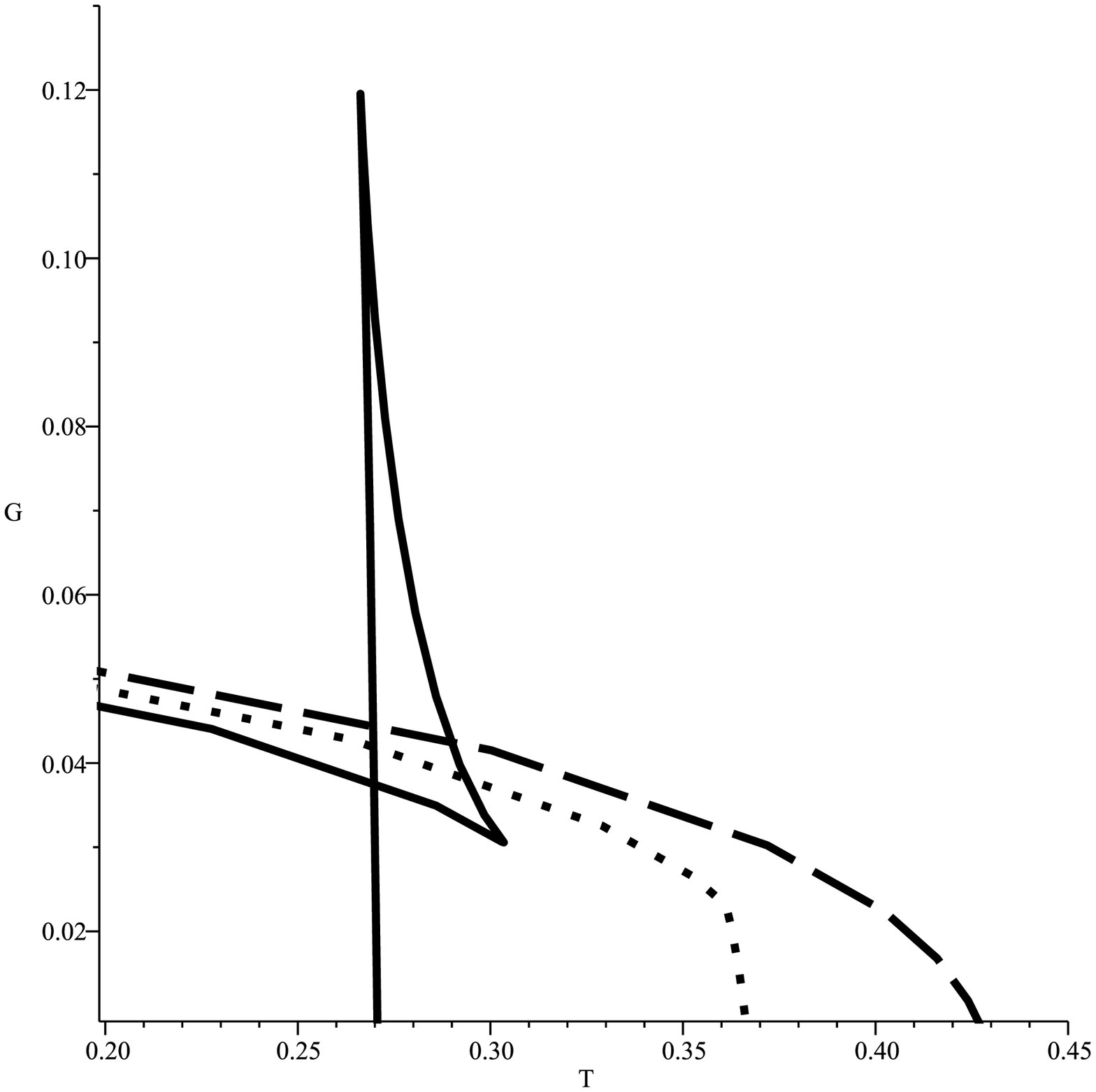}%
\end{array}
$%
\caption{ $P-v$ (left), $T-v$ (middle) and $G-T$ (right) diagrams in GB
gravity for $\protect\alpha=0.1$, $q=1$ and $\protect\beta=1.5$ ($d=7$).
\newline
$P-v$ diagram, from up to bottom $T=1.2T_{c}$, $T=1.1T_{c}$, $T=T_{c}$, $%
T=0.85T_{c}$ and $T=0.75T_{c}$, respectively. \newline
$T-v$ diagram, from up to bottom $P=1.2P_{c}$, $P=1.1P_{c}$. $P=P_{c}$, $%
P=0.85P_{c}$ and $P=0.75P_{c}$, respectively. \newline $G-T$
diagram for $P=0.5P_{c}$ (continuous line), $P=P_{c}$ (dotted
line) and $P=1.5P_{c}$ (dashed line). } \label{Fig3}
\end{figure}
\begin{figure}[tbp]
$%
\begin{array}{ccc}
\epsfxsize=5cm \epsffile{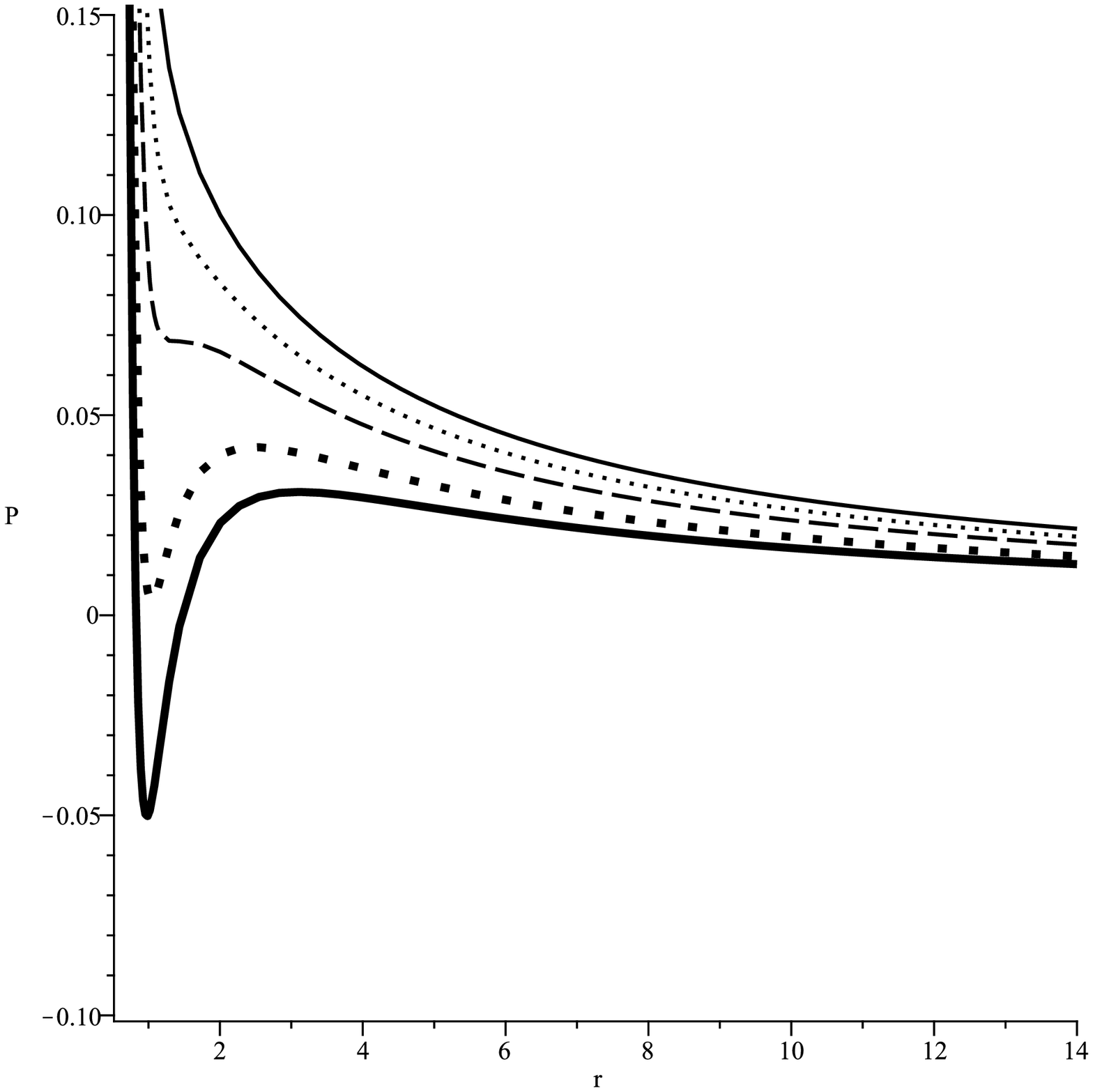} & \epsfxsize=5cm %
\epsffile{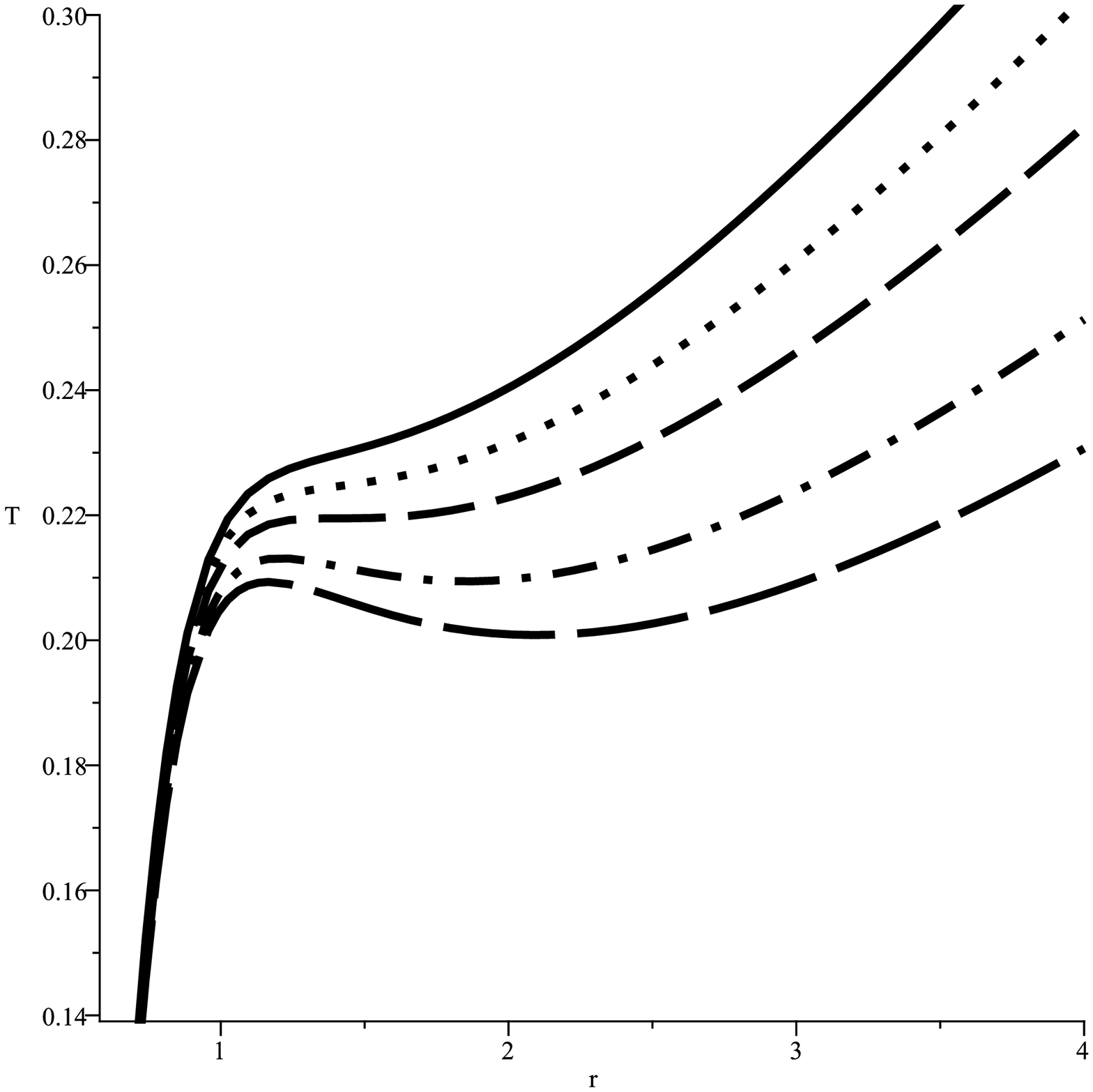} & \epsfxsize=5cm %
\epsffile{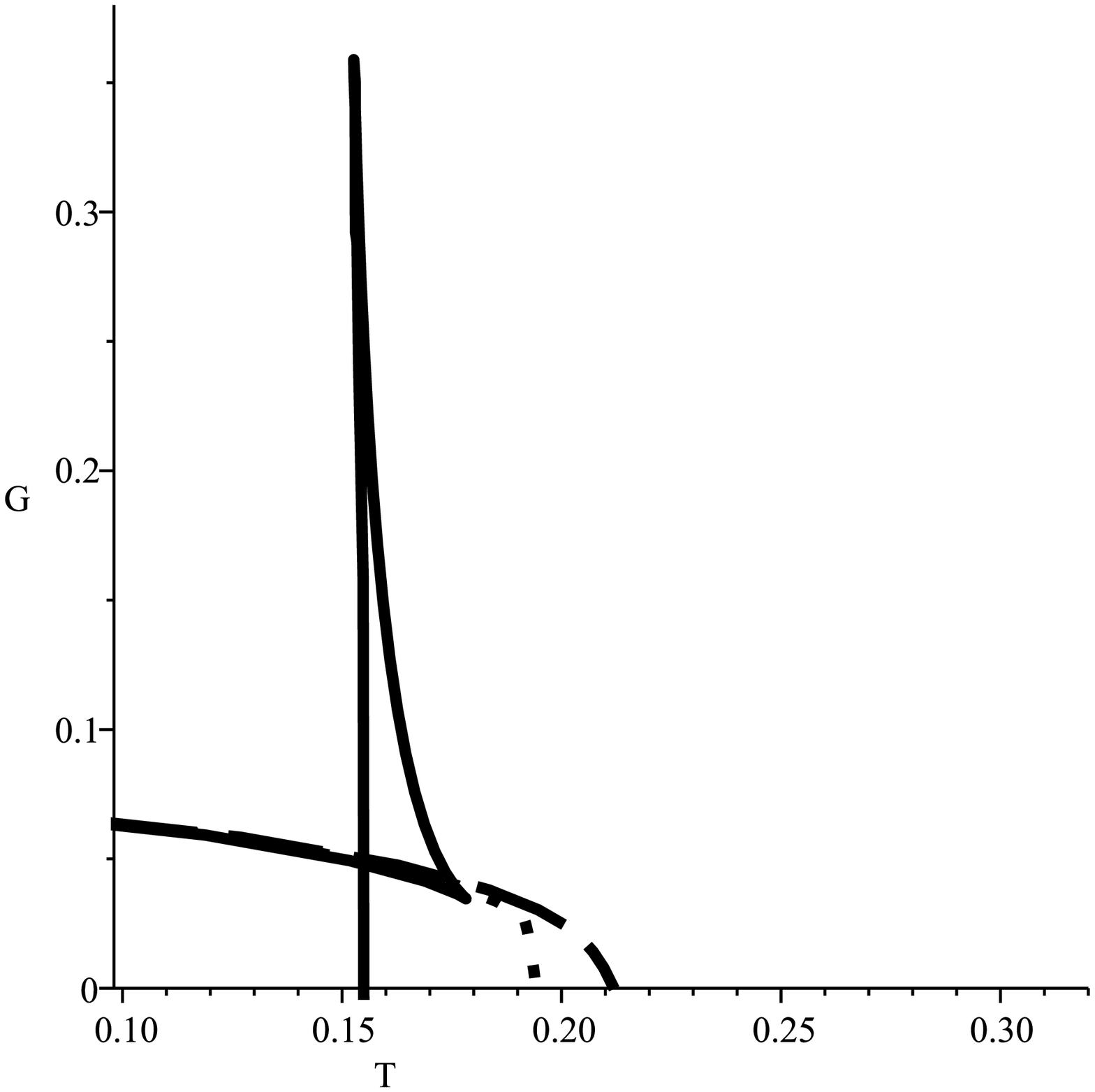}%
\end{array}
$%
\caption{ $P-v$ (left), $T-v$ (middle) and $G-T$ (right) diagrams in GB
gravity for $\protect\beta=1$, $q=1$ and $\protect\alpha=0.5$ ($d=7$).
\newline
$P-v$ diagram, from up to bottom $T=1.2T_{c}$, $T=1.1T_{c}$, $T=T_{c}$, $%
T=0.85T_{c}$ and $T=0.75T_{c}$, respectively. \newline
$T-v$ diagram, from up to bottom $P=1.2P_{c}$, $P=1.1P_{c}$. $P=P_{c}$, $%
P=0.85P_{c}$ and $P=0.75P_{c}$, respectively. \newline $G-T$
diagram for $P=0.5P_{c}$ (continuous line), $P=P_{c}$ (dotted
line) and $P=1.5P_{c}$ (dashed line). } \label{Fig4}
\end{figure}
\begin{figure}[tbp]
$%
\begin{array}{ccc}
\epsfxsize=5cm \epsffile{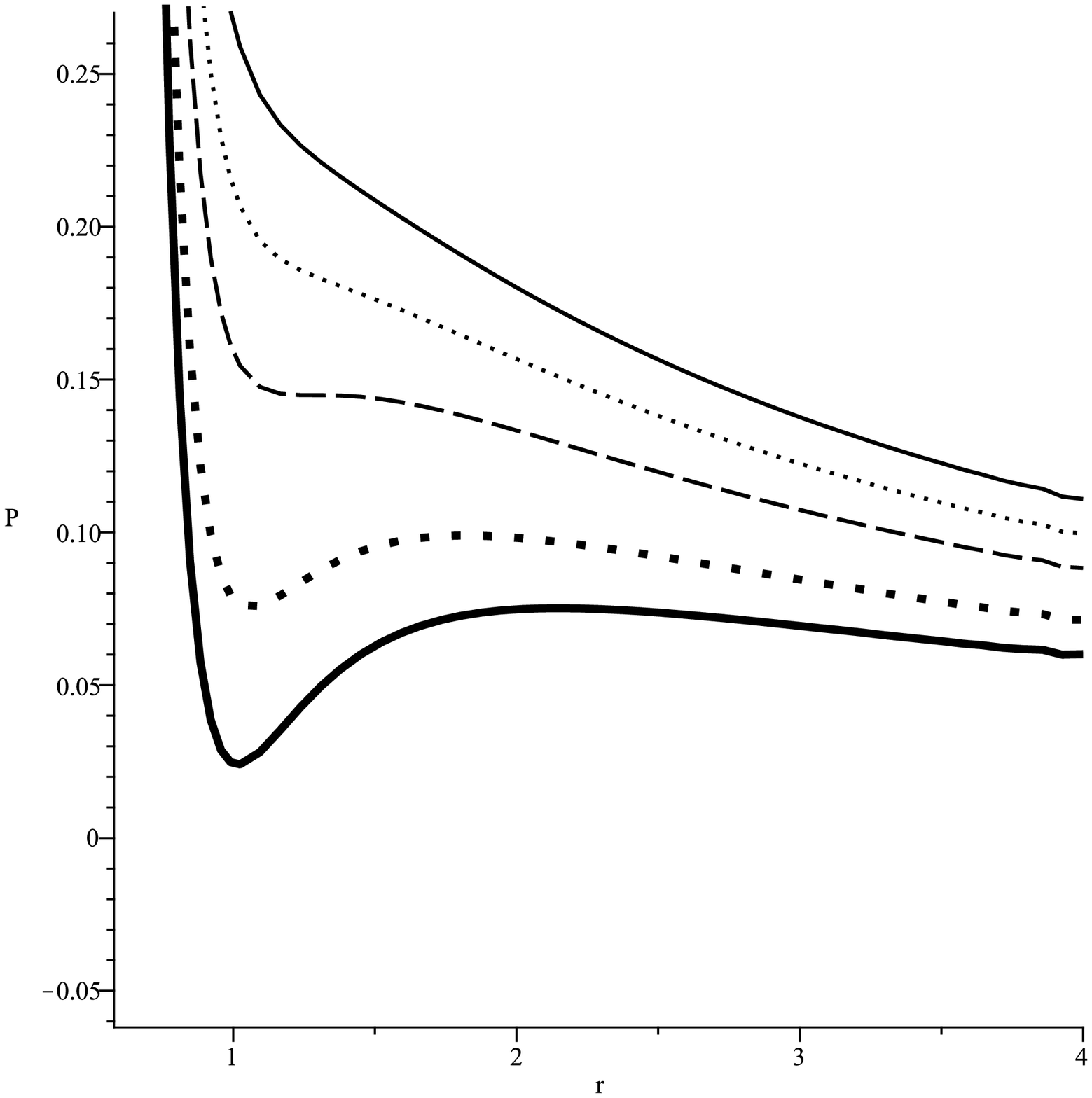} & \epsfxsize=5cm %
\epsffile{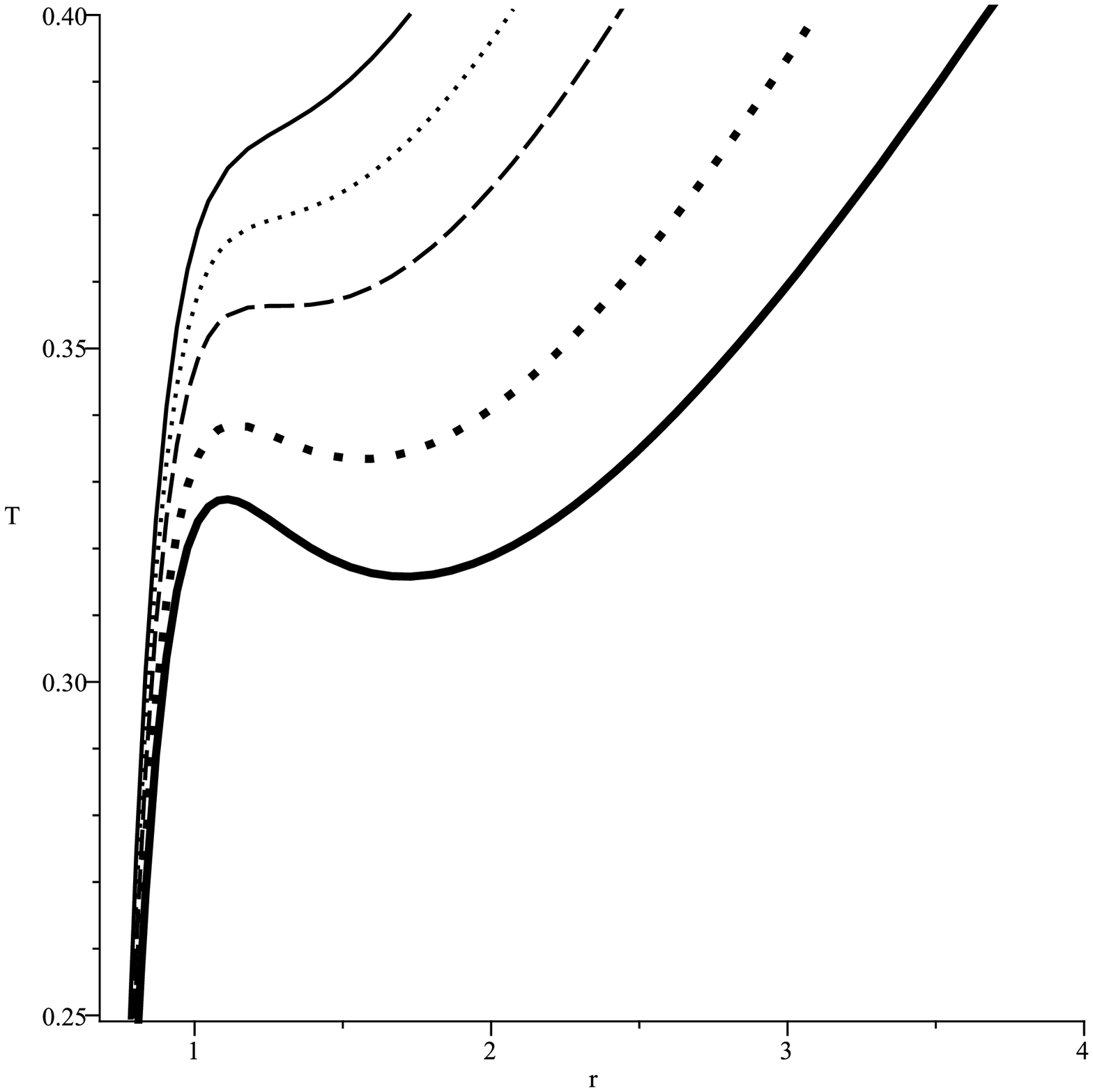} & \epsfxsize=5cm %
\epsffile{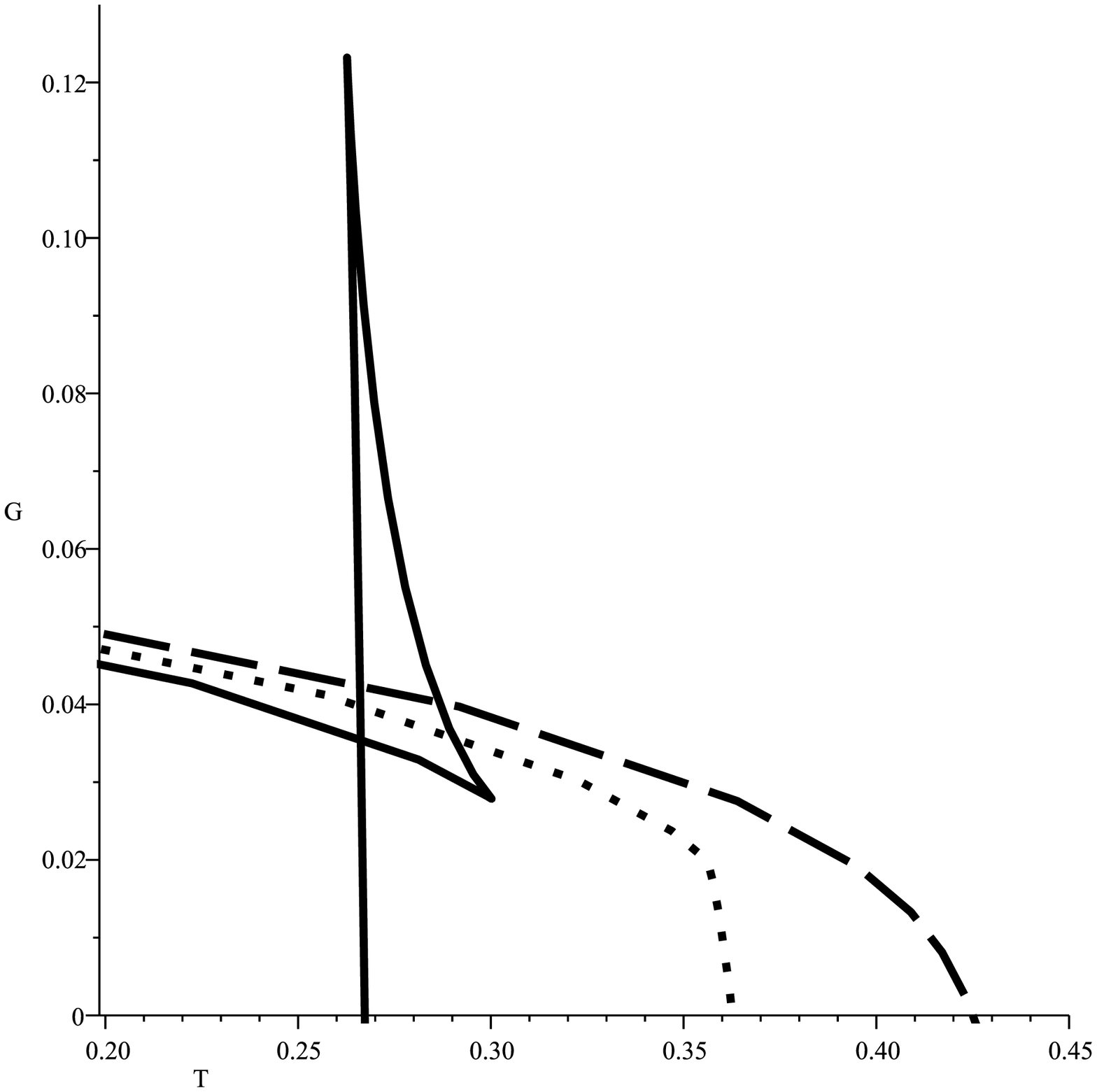}%
\end{array}
$%
\caption{ $P-v$ (left), $T-v$ (middle) and $G-T$ (right) diagrams
in TOL gravity for $\protect\alpha =0.1$, $q=1$ and $\protect\beta
=1.5$ ($d=7$).
\newline
$P-v$ diagram, from up to bottom $T=1.2T_{c}$, $T=1.1T_{c}$, $T=T_{c}$, $%
T=0.85T_{c}$ and $T=0.75T_{c}$, respectively. \newline
$T-v$ diagram, from up to bottom $P=1.2P_{c}$, $P=1.1P_{c}$. $P=P_{c}$, $%
P=0.85P_{c}$ and $P=0.75P_{c}$, respectively. \newline $G-T$
diagram for $P=0.5P_{c}$ (continuous line), $P=P_{c}$ (dotted
line) and $P=1.5P_{c}$ (dashed line). } \label{Fig6}
\end{figure}
\begin{figure}[tbp]
$%
\begin{array}{ccc}
\epsfxsize=5cm \epsffile{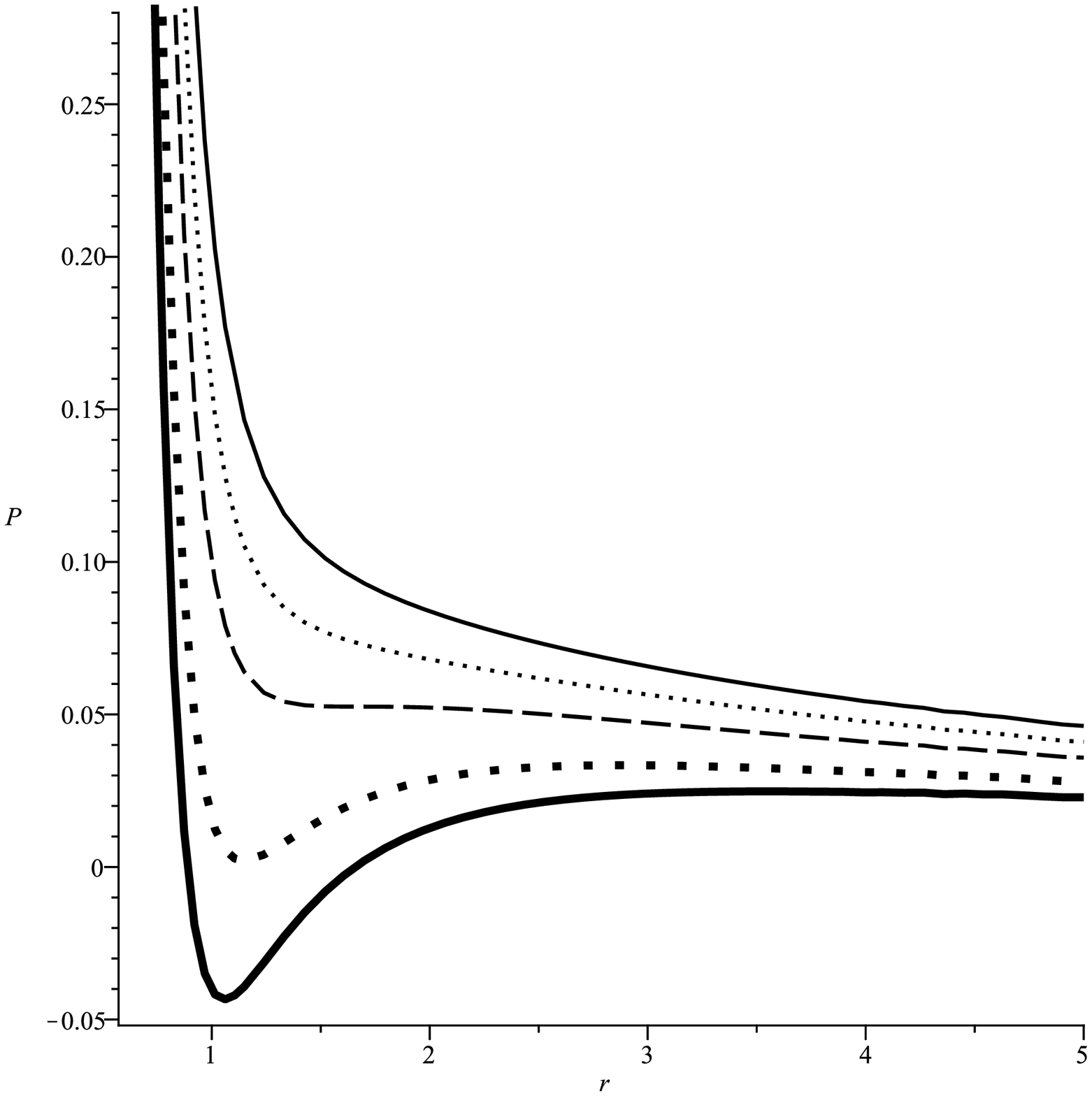} & \epsfxsize=5cm %
\epsffile{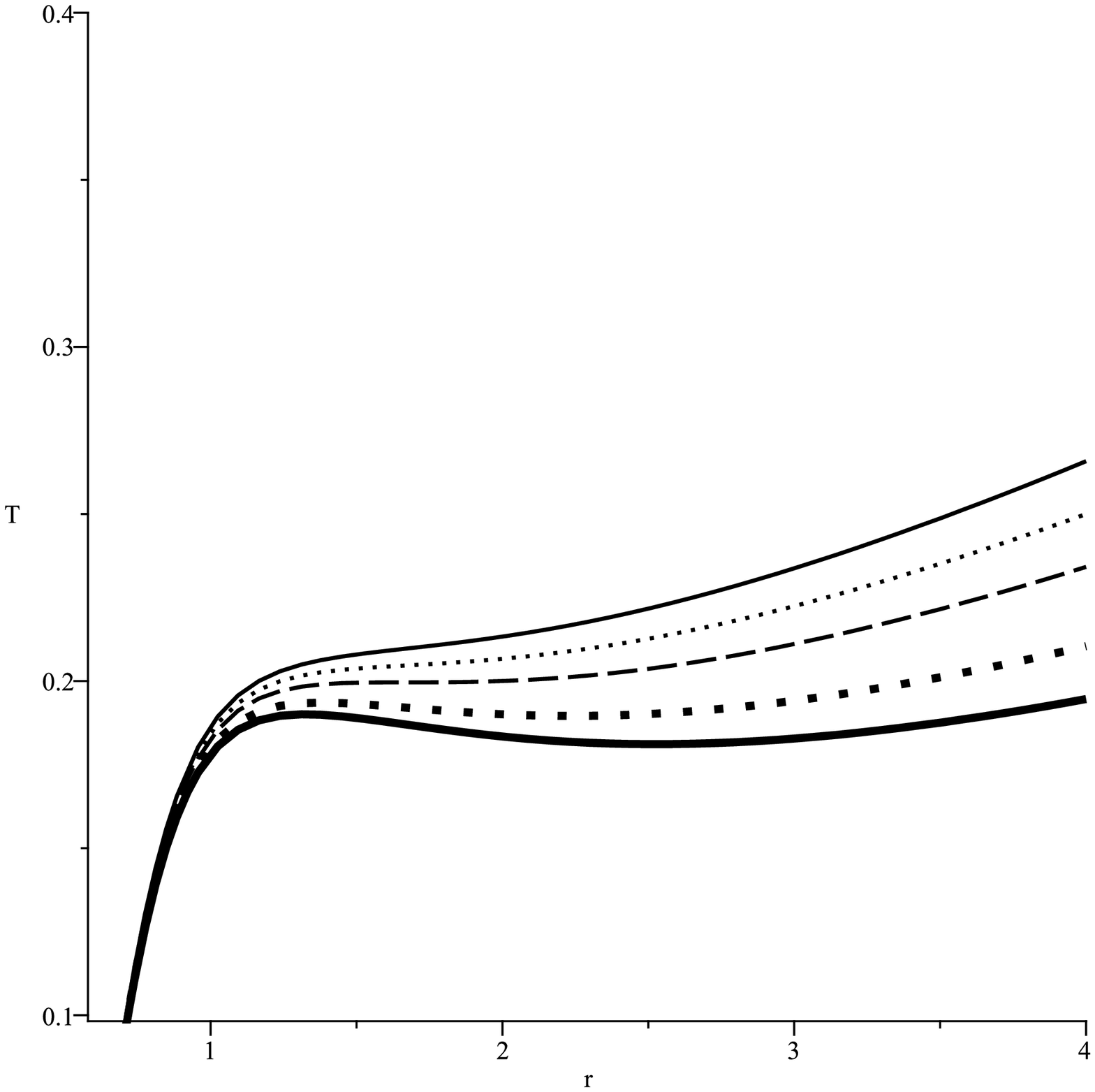} & \epsfxsize=5cm %
\epsffile{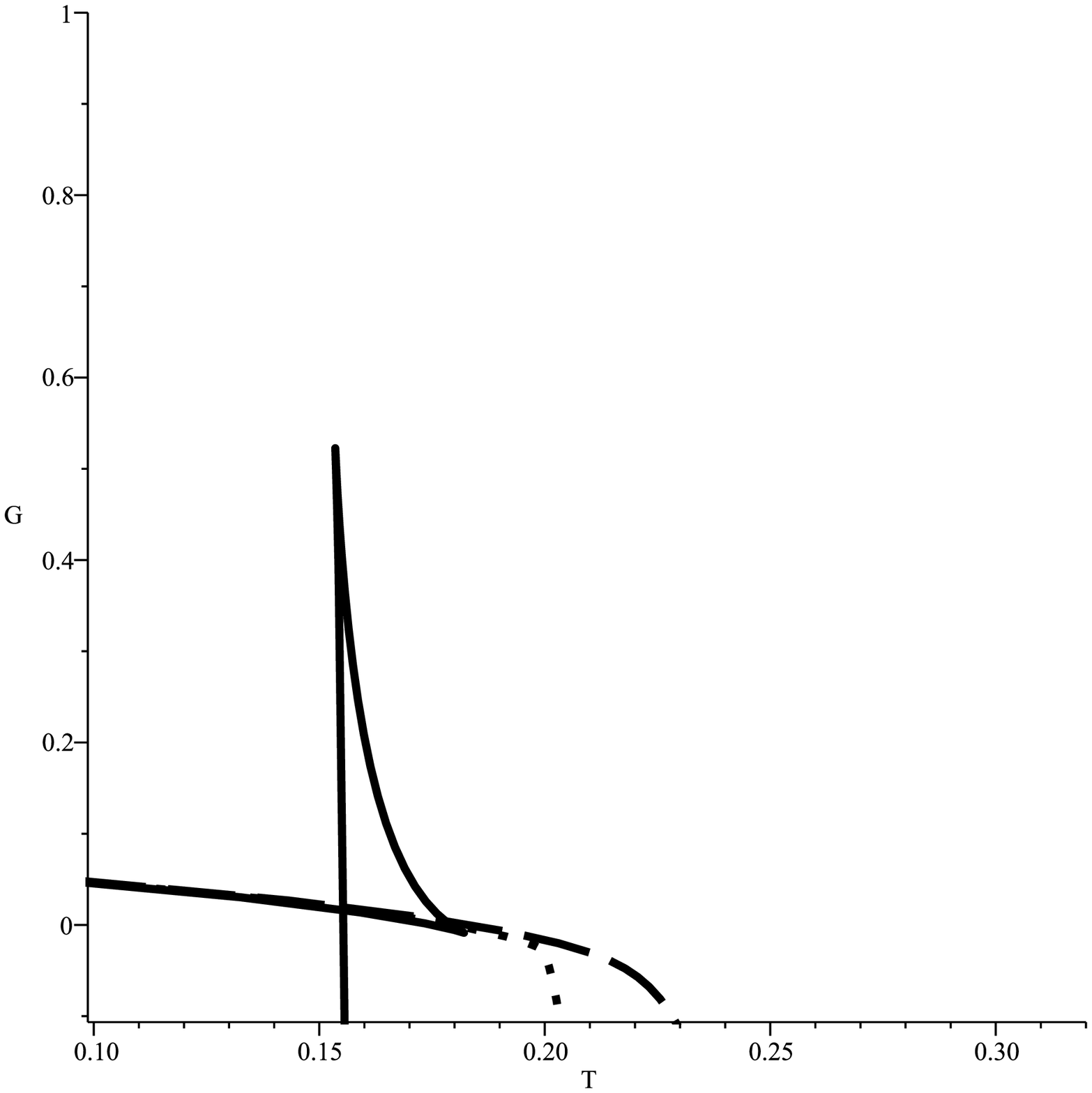}%
\end{array}
$%
\caption{ $P-v$ (left), $T-v$ (middle) and $G-T$ (right) diagrams
in TOL gravity for $\protect\beta =1$, $q=1$, and $\protect\alpha
=0.5$ ($d=7$).
\newline
$P-v$ diagram, from up to bottom $T=1.2T_{c}$, $T=1.1T_{c}$, $T=T_{c}$, $%
T=0.85T_{c}$ and $T=0.75T_{c}$, respectively. \newline
$T-v$ diagram, from up to bottom $P=1.2P_{c}$, $P=1.1P_{c}$. $P=P_{c}$, $%
P=0.85P_{c}$ and $P=0.75P_{c}$, respectively. \newline $G-T$
diagram for $P=0.5P_{c}$ (continuous line), $P=P_{c}$ (dotted
line) and $P=1.5P_{c}$ (dashed line). } \label{Fig7}
\end{figure}

\section{Discussion on the results of diagrams}

As one can confirm, higher orders of Lovelock gravity modify the
phase diagrams and critical values of volume, pressure and
temperature (see tables $1-6$\ and related Figs. of \ref{Fig1} -
\ref{Fig7}). It is clear that considering higher orders of
Lovelock gravity leads to different kinds of thermodynamical
systems which was also evident through calculated conserved
quantities.

In this paper, we consider thermodynamical effects of Lovelock
gravity up to GB and TOL, separately. It is evident that critical
temperature and pressure are decreasing functions of $\alpha$ and
also orders of Lovelock gravity whereas the critical volume is an
increasing function of these two factors (see Figs. \ref{Fig8},
\ref{Fig9}, \ref{Fig11} and \ref{Fig12}). On the other hand, the
energy gap between two phases increases drastically by
transforming from lower order of Lovelock gravity to higher one
(see Figs. \ref{Fig8}, \ref{Fig9}, \ref{Fig11} and \ref{Fig12}
right panels) or by increasing $\alpha$ for each order (see Figs.
\ref{Fig8} and \ref{Fig9} right panels). In addition, the length
of subcritical isobars increases which means that the
single phase region of small/large black holes decreases (see Figs. \ref%
{Fig9} and \ref{Fig12} middle panels). Also, phase transitions
region is an increasing function of $\alpha$ and order of Lovelock
gravity (see Figs. \ref{Fig9} and \ref{Fig12} left panels).

To conclude, plotted figures show that the highest critical
temperature and pressure, and the lowest critical volume and
energy gap belong to Einstein gravity. On the other hand, system
needs higher temperature to have phase transition in Einstein
gravity. Whereas the critical temperature decreases by considering
higher orders of Lovelock gravity or increasing Lovelock
coefficients. Considering the fact that increasing Lovelock
parameters and/or adding higher orders of Lovelock gravity
increase the power of gravity, one may say that in stronger
gravitational regimes, phase transitions take place in lower
temperature.

\begin{figure}[tbp]
$%
\begin{array}{ccc}
\epsfxsize=5cm \epsffile{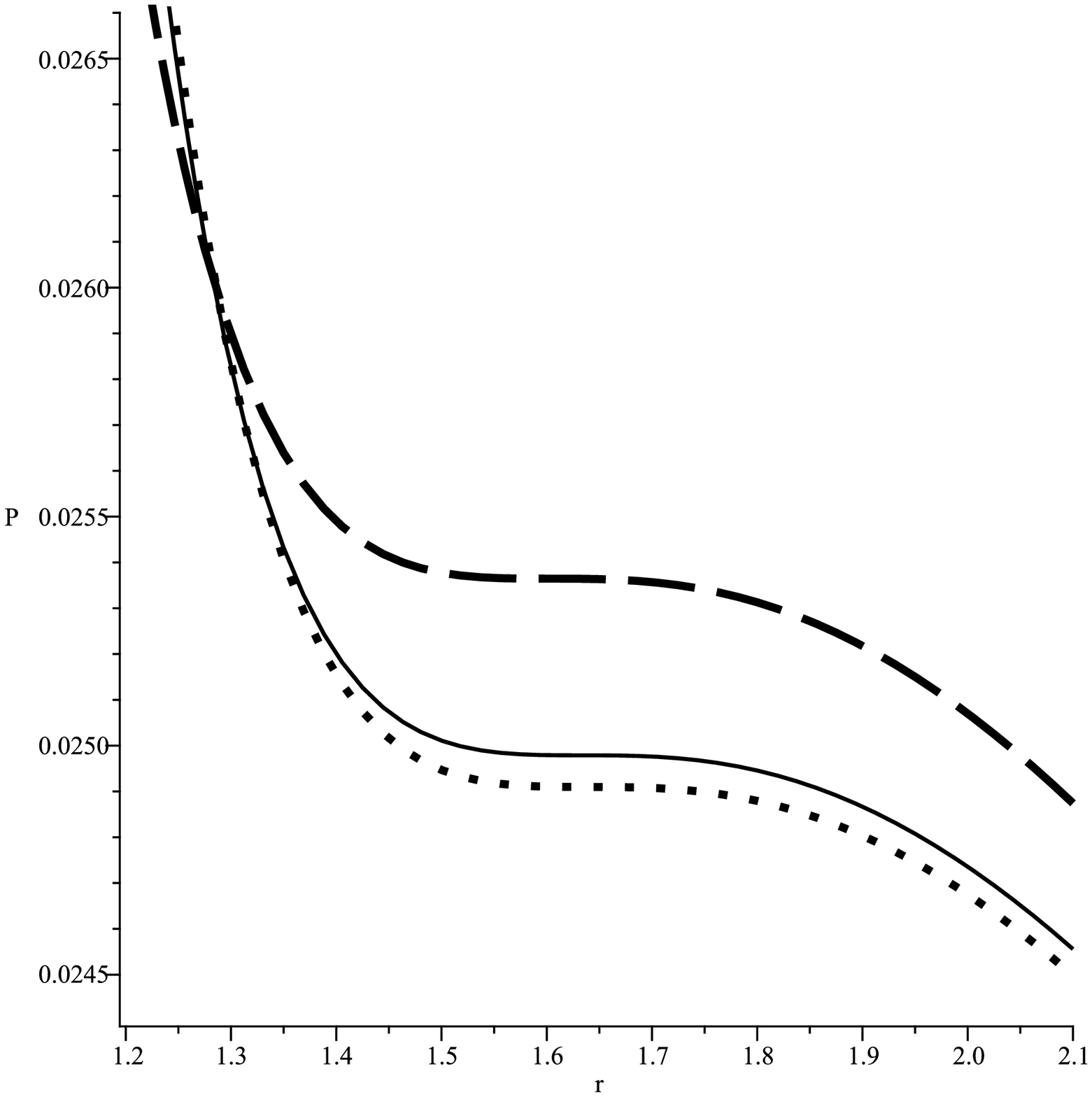} & \epsfxsize=5cm %
\epsffile{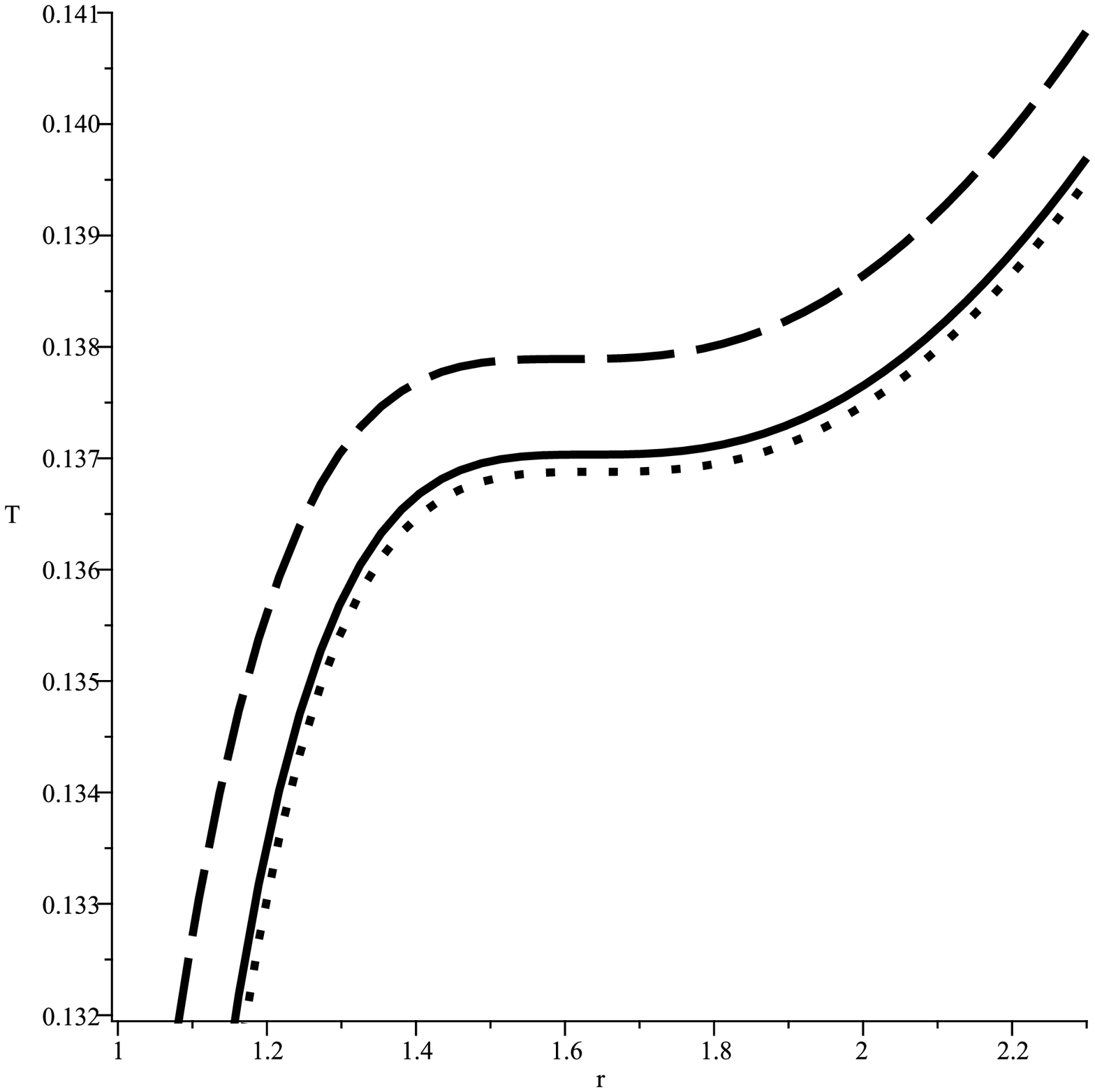} & \epsfxsize=5cm %
\epsffile{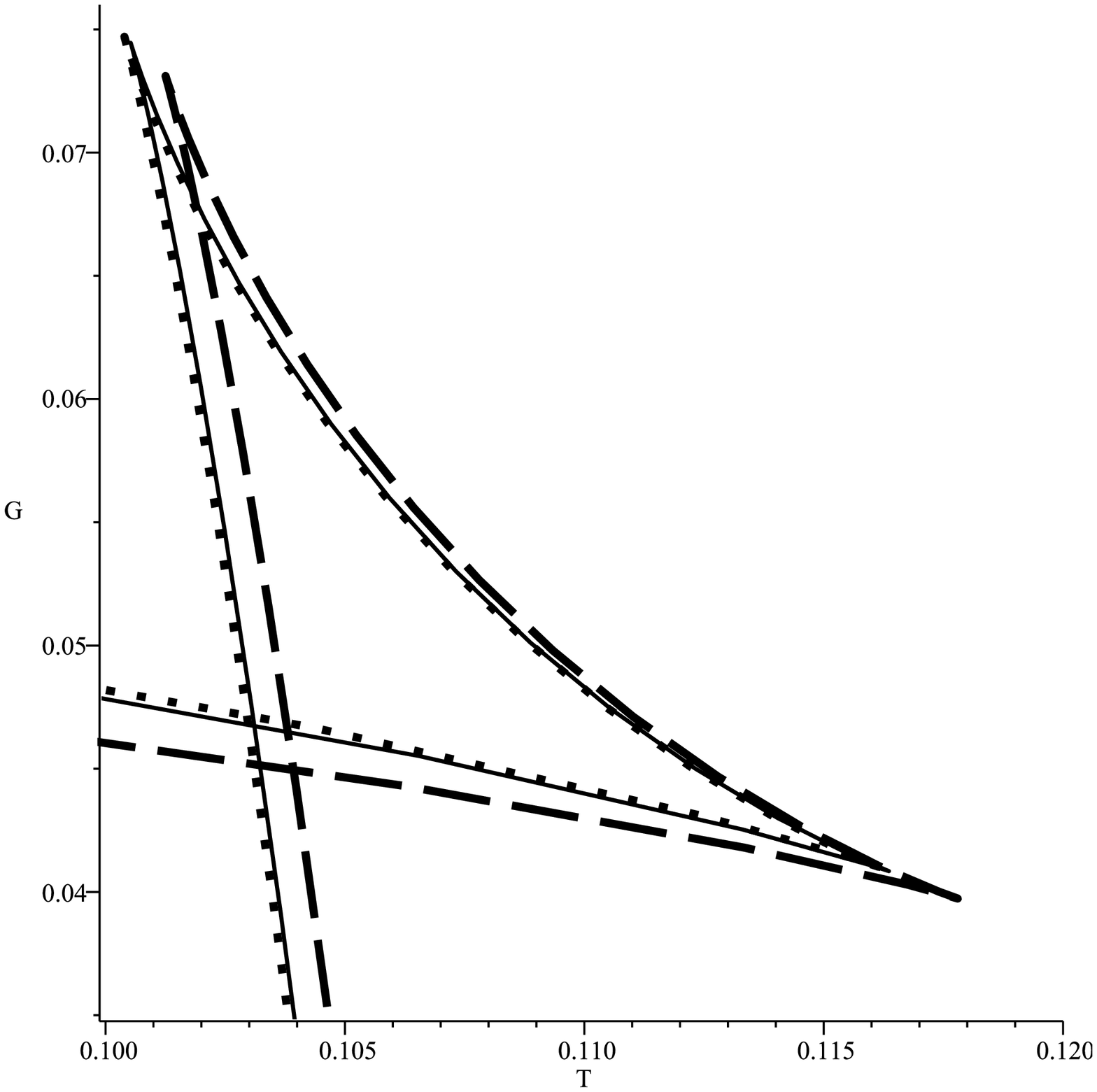}%
\end{array}
$%
\caption{ $P-v$ (left), $T-v$ (middle) and $G-T$ (right) diagrams
in GB gravity for $\protect\alpha =0.1$, $q=1$, $d=5$,
$\protect\beta =0.5$ (dashed line), $\protect\beta =1$ (continuous
line) and $\protect\beta =1.5$ (dotted line). \newline $P-v$
diagram for $T=T_{c}$, $T-v$ diagram for $P=P_{c}$ and $G-T$
diagram for $P=0.5P_{c}$.} \label{Fig5}
\end{figure}

\begin{figure}[tbp]
$%
\begin{array}{ccc}
\epsfxsize=5cm \epsffile{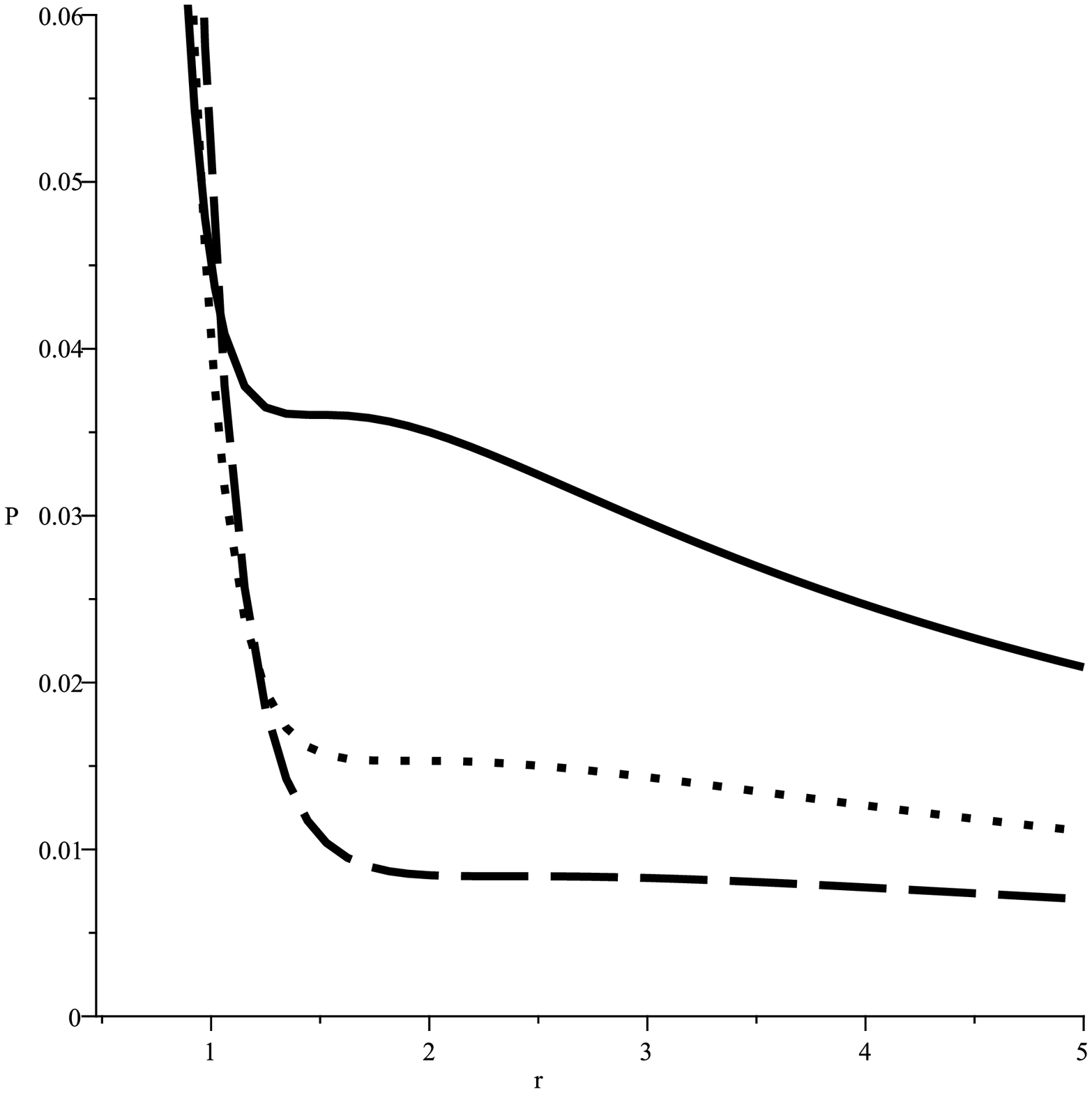} & \epsfxsize%
=5cm \epsffile{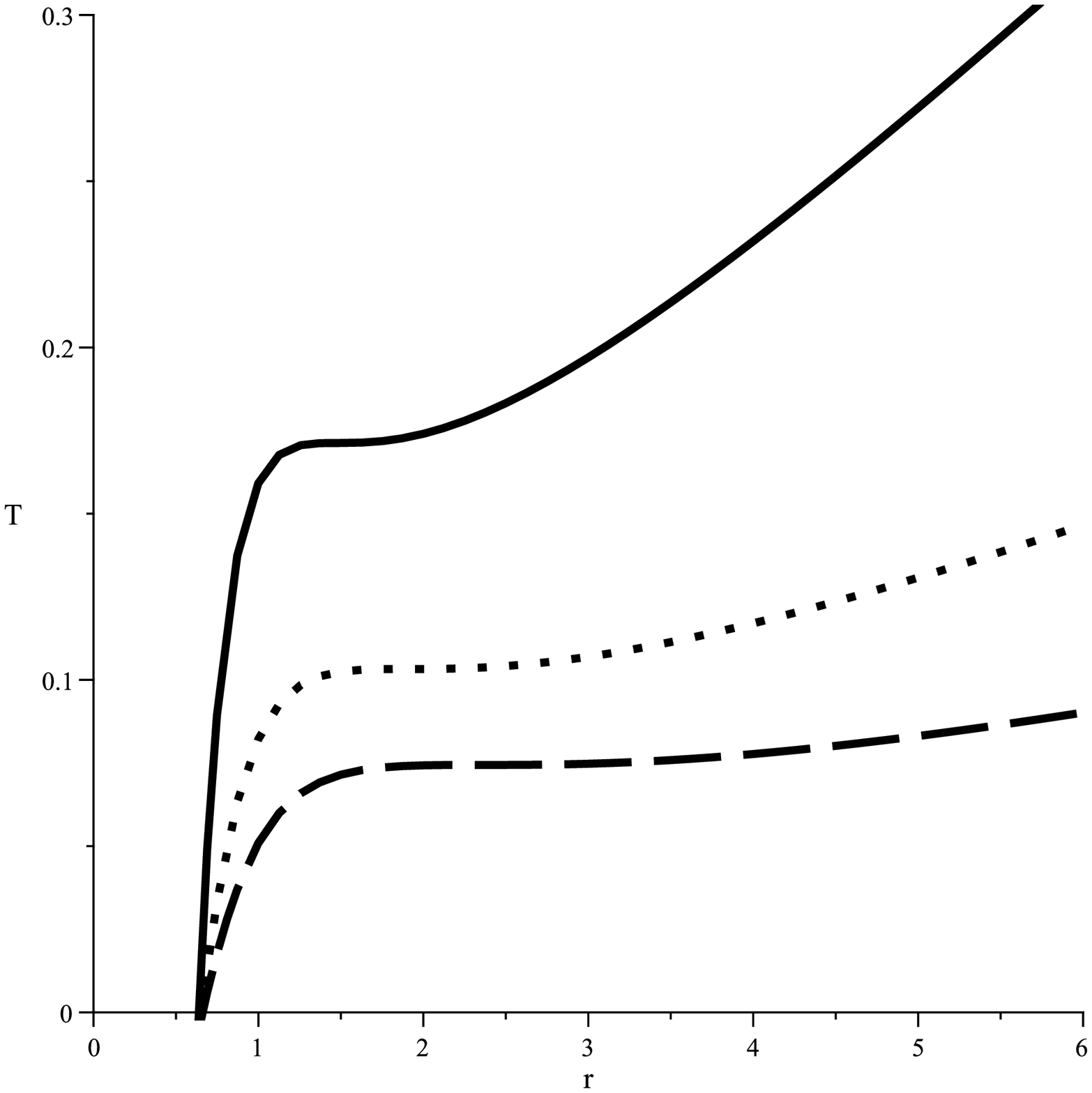} & \epsfxsize=5cm %
\epsffile{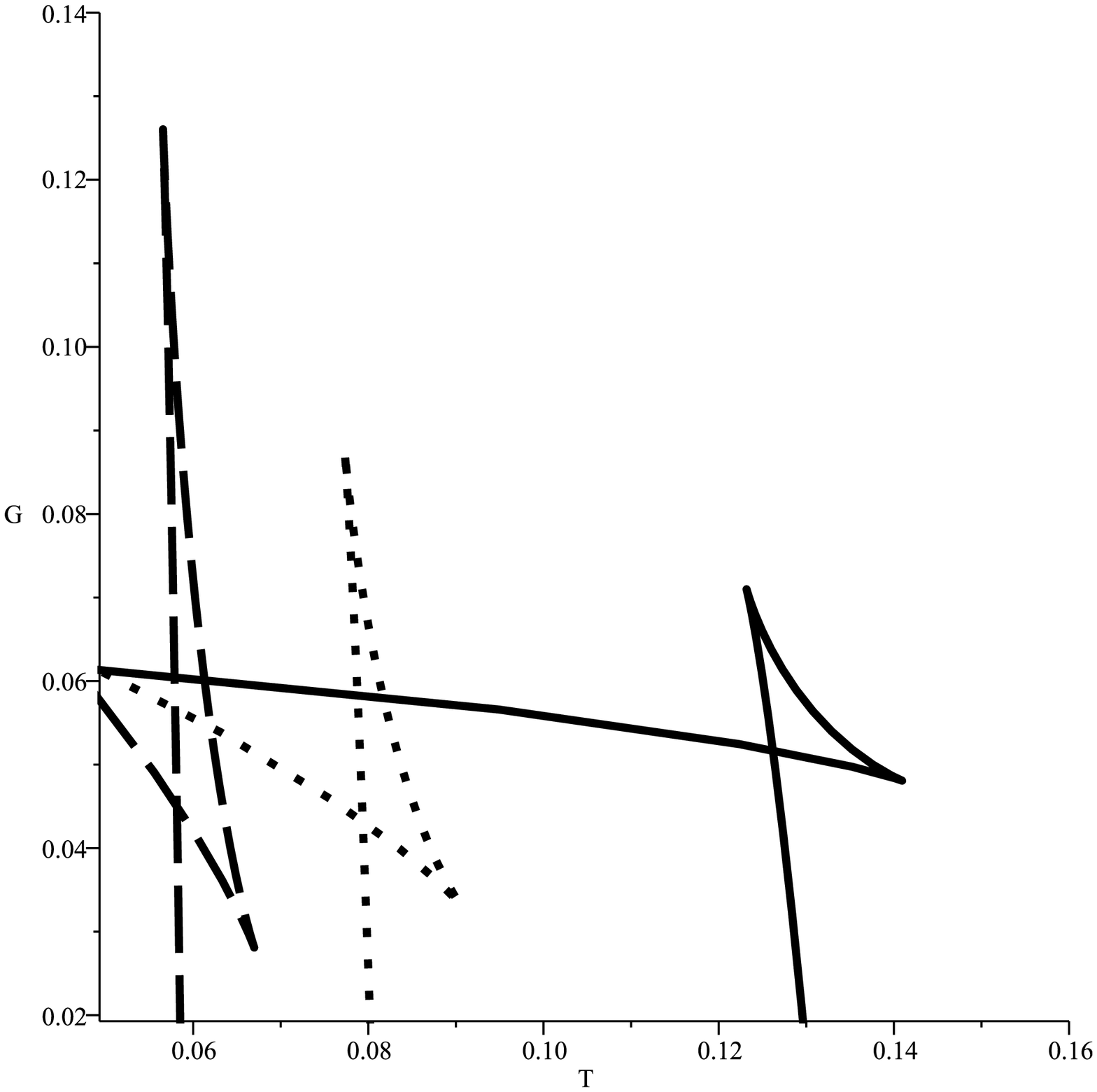}%
\end{array}
$%
\caption{ $P-v$ (left), $T-v$ (middle) and $G-T$ (right) diagrams in GB
gravity for $\protect\beta =1$, $q=1$, $d=5$, $\protect\alpha =0$
(continuous line), $\protect\alpha =0.3$ (dotted line) and $\protect\alpha %
=0.7$ (dashed line). \newline
$P-v$ diagram for $T=T_{c}$, $T-v$ diagram for $P=P_{c}$, $G-T$ diagram for $%
P=0.5P_{c}$.} \label{Fig8}
\end{figure}

\begin{figure}[tbp]
$%
\begin{array}{ccc}
\epsfxsize=5cm \epsffile{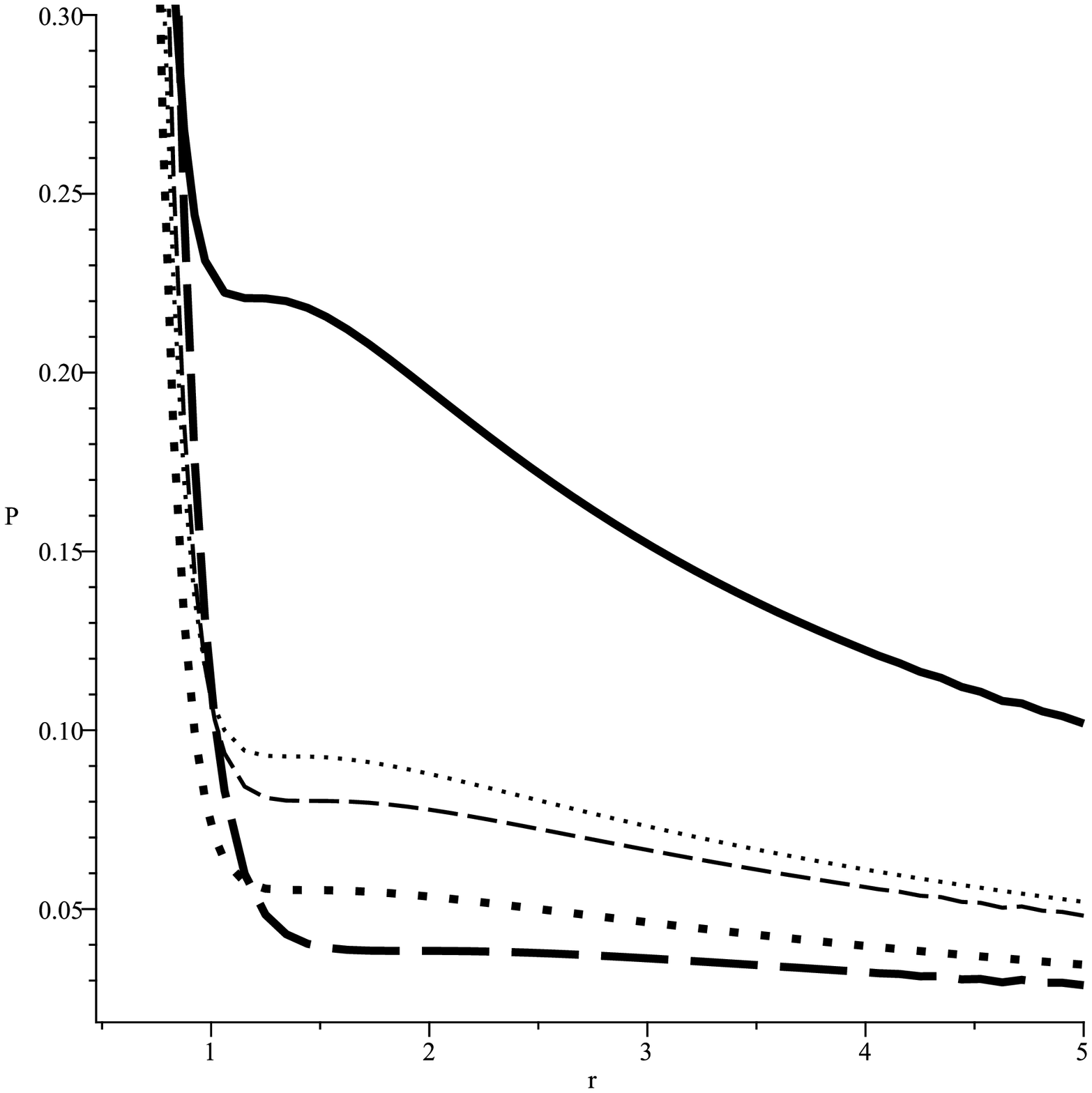} & \epsfxsize%
=5cm \epsffile{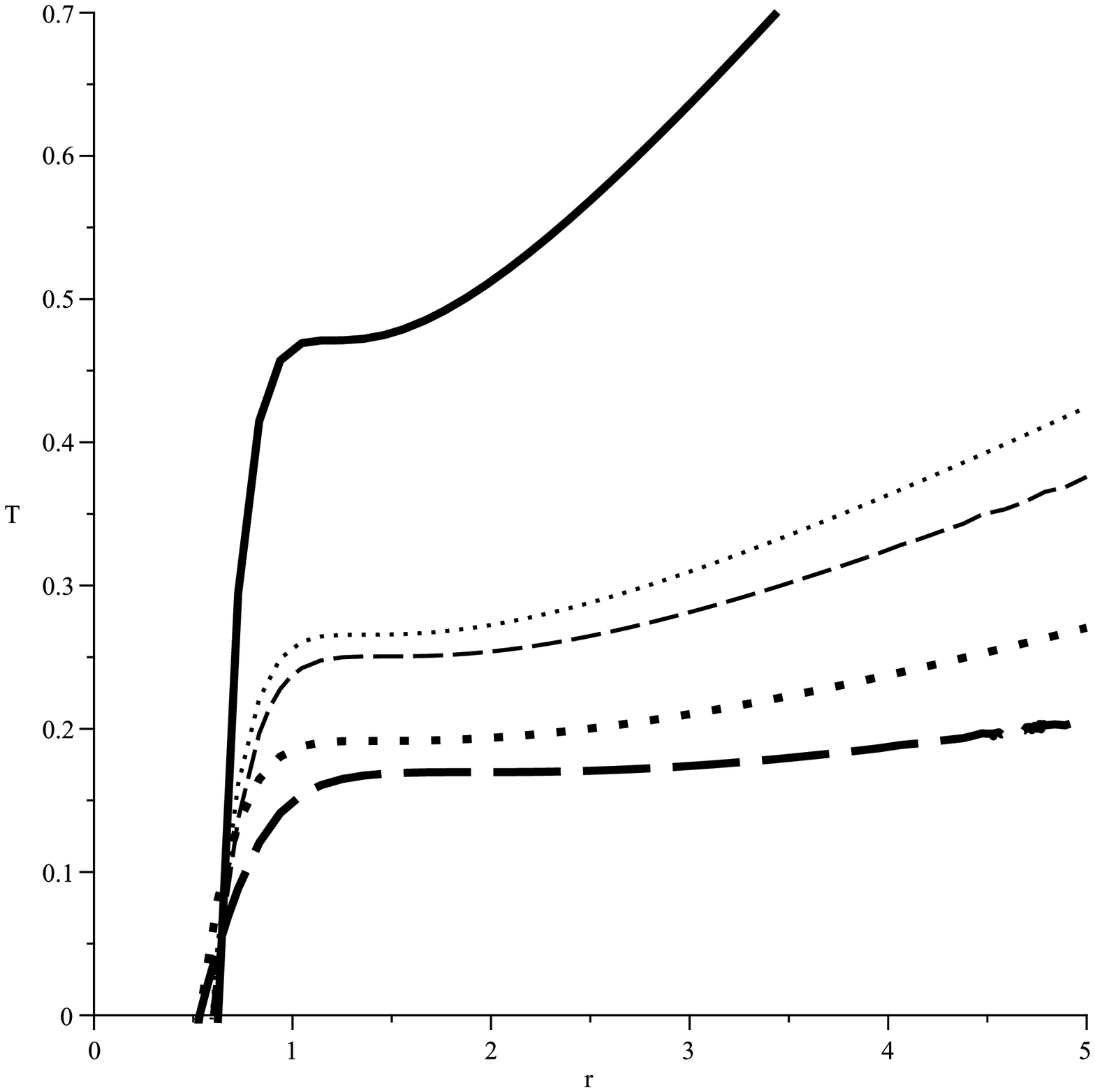} & \epsfxsize=5cm %
\epsffile{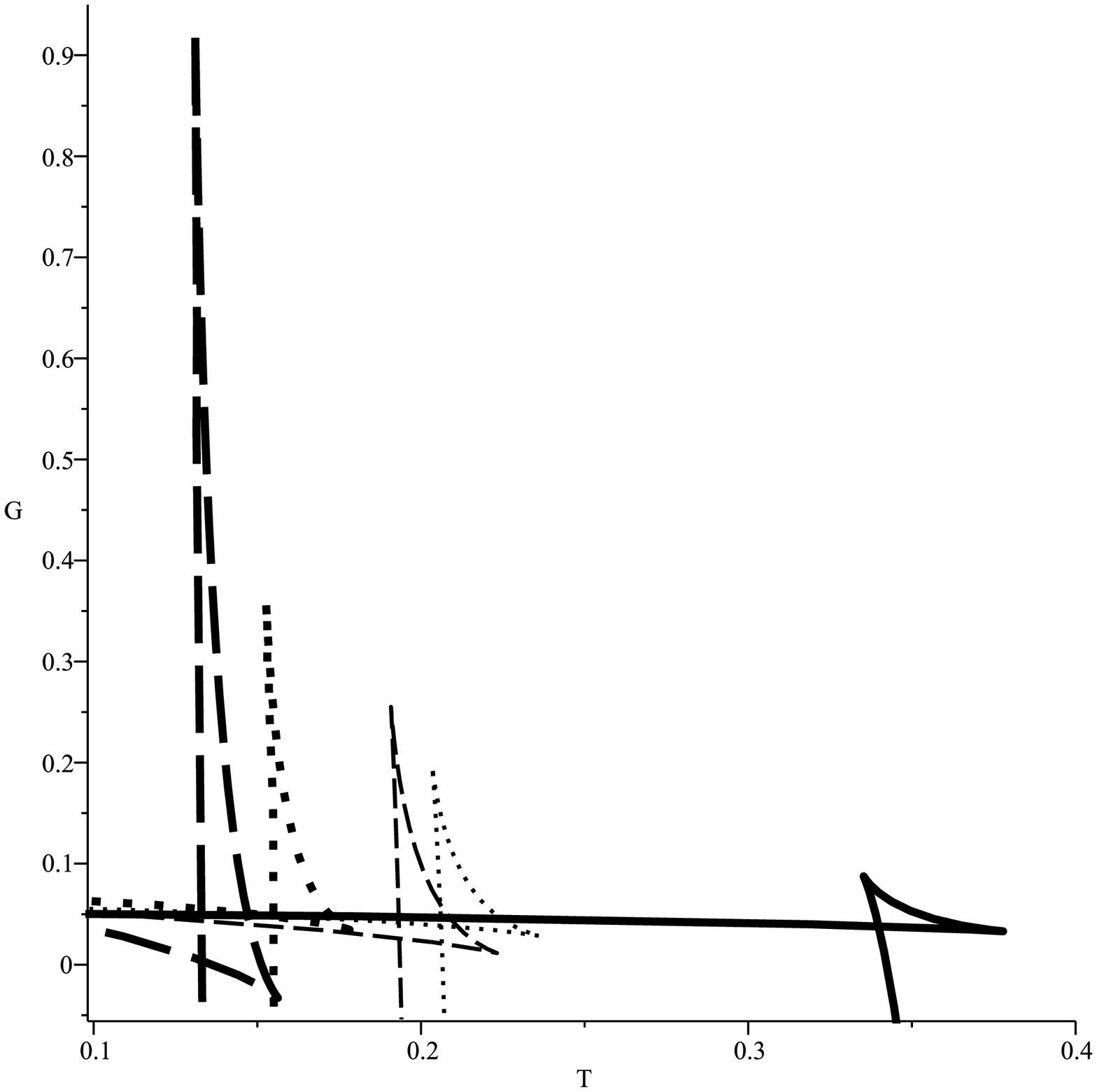}%
\end{array}
$%
\caption{$P-v$ diagram for $T=T_{c}$ (left), $T-v$ diagram for
$P=P_{c}$ (middle) and $G-T$ diagram for $P=0.5P_{c}$ (right) with
$\protect\beta =1$, $q=1$ and $d=7$. \newline continuous line (EN
gravity or $\alpha=0$), dotted line (GB gravity with $\alpha
=0.3$), dotted-bold line (GB gravity with $\alpha =0.7$), dashed
line (TOL gravity with $\alpha =0.3$) and dashed-bold line (TOL
gravity with $\alpha =0.7$)} \label{Fig9}
\end{figure}

As for the effects of NED, we find the following results. It is
evident that the critical temperature in which swallow tail is
formed (see Figs. \ref{Fig5}, \ref{Fig11} and \ref{Fig12} right
panels) and critical pressure (see Figs. \ref{Fig5}, \ref{Fig11}
and \ref{Fig12} left panels) are decreasing functions of
nonlinearity parameter whereas the critical volume is an
increasing function. In addition, the length of subcritical
isobars is an increasing function of nonlinearity parameter. This
means that single phase region of small/large black holes is a
decreasing function of $\beta$ (see Figs. \ref{Fig5}, \ref{Fig11}
and \ref{Fig12} middle panels). Therefore, for higher values of
nonlinearity parameter (weak nonlinearity strength), black holes
need to absorb less mass in order to have phase transition.
Another issue that must be taken into account is the fact that as
$\beta$ increases, the gap between isobars decreases whereas for
small $\beta$ this gap is greater.

Due to the fact that we take into account BI type models, for
large values of nonlinearity parameter, they will lead to Maxwell
theory. Obtained results show that the lowest critical temperature
and pressure and highest critical volume belong to Maxwell theory.
On the other hand, one can conclude that the power of the
nonlinearity causes the system to need higher critical temperature
to have a phase transition. This effect is opposite of what was
observed for gravity.

\begin{figure}[tbp]
$%
\begin{array}{ccc}
\epsfxsize=5cm \epsffile{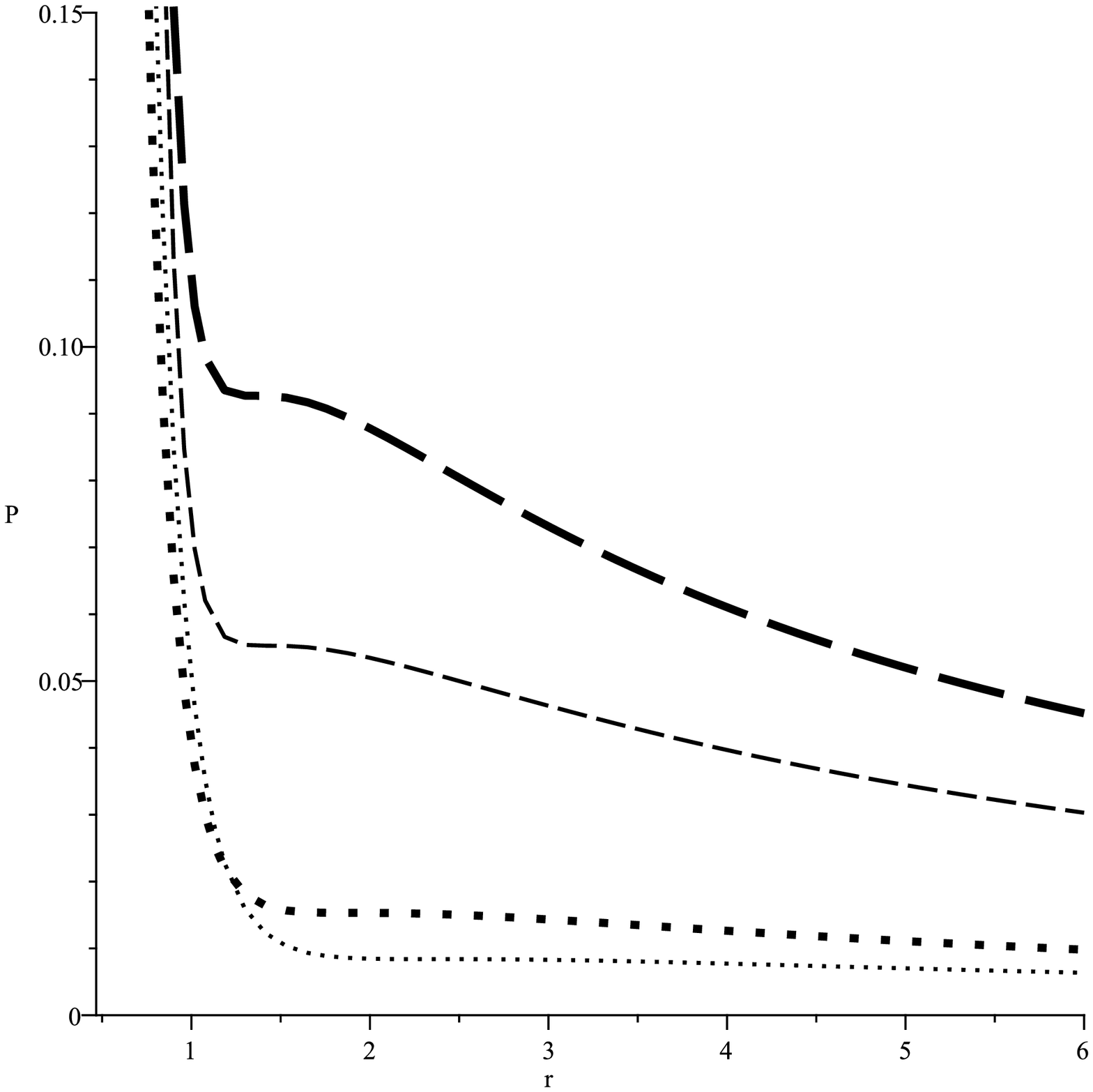} & \epsfxsize=5cm %
\epsffile{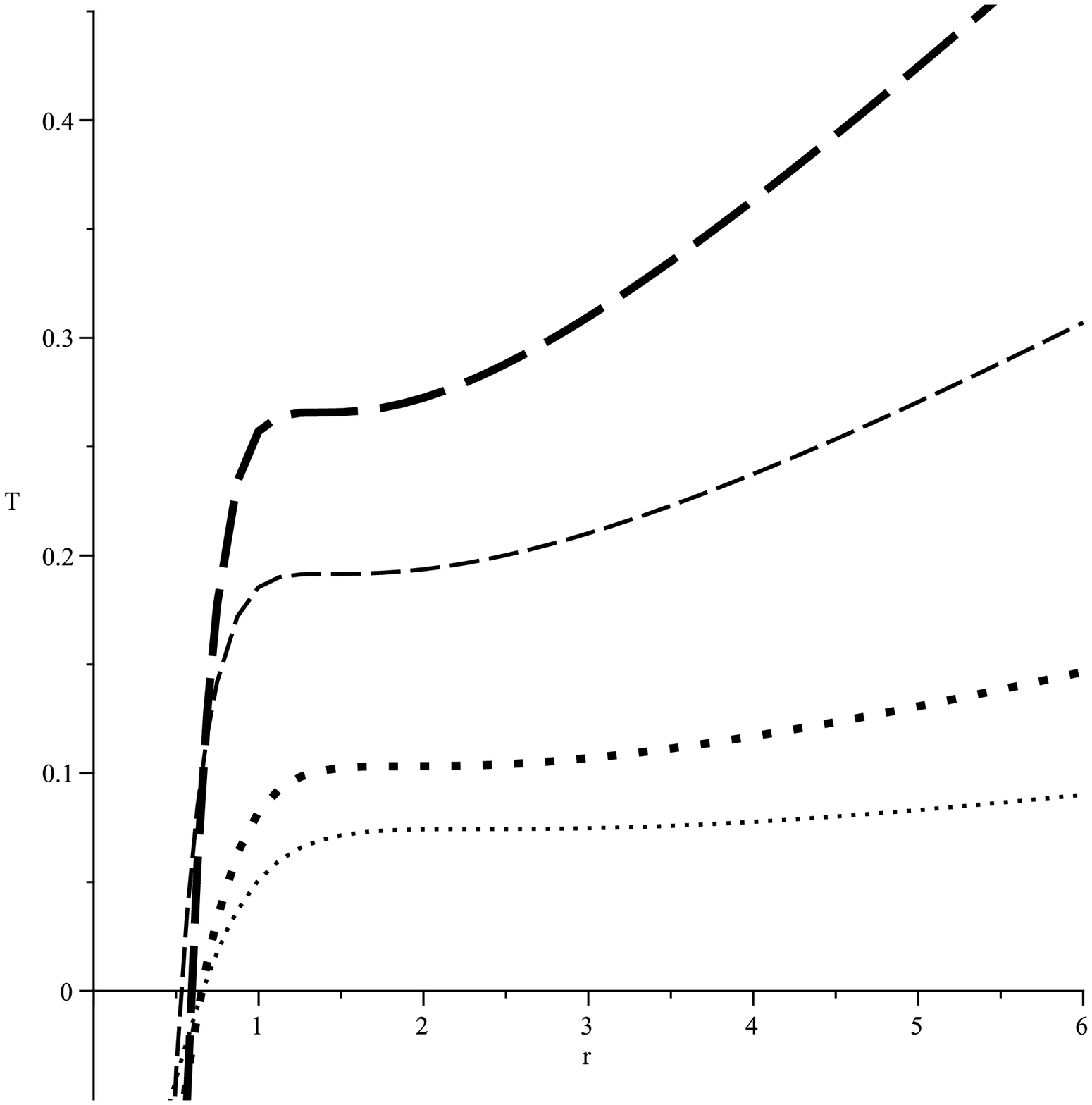} & \epsfxsize=5cm %
\epsffile{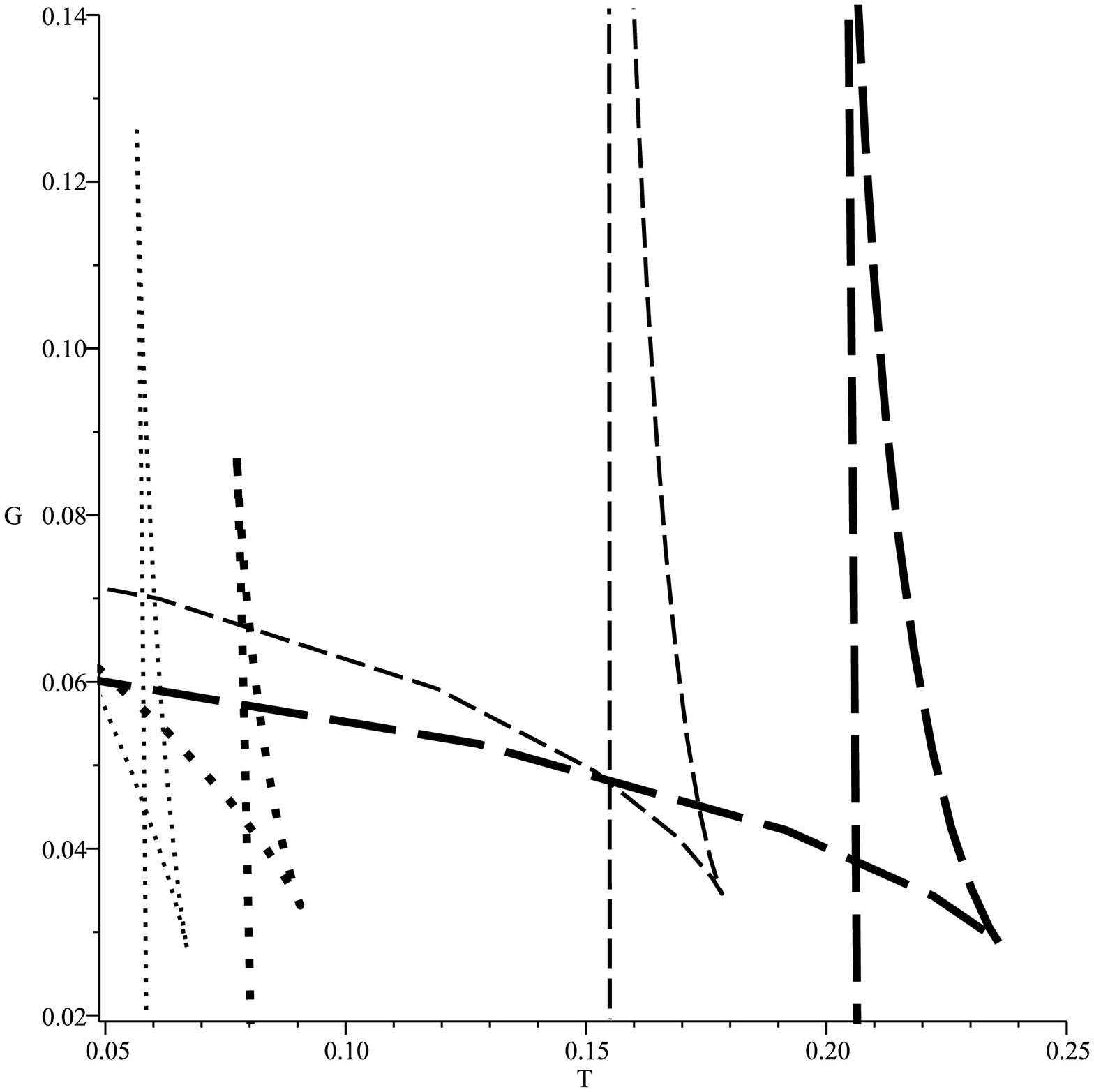}%
\end{array}
$%
\caption{$P-v$ diagram for $T=T_{c}$ (left), $T-v$ diagram for
$P=P_{c}$ (middle)and $G-T$ diagram for $P=0.5P_{c}$ (right) with
$\beta=1$ and $q=1$ for GB gravity. \newline dotted line ($d=5$
and $\alpha =0.7$), dotted-bold line ($d=5$ and $\alpha =0.3$),
dashed line ($d=7$ and $\alpha =0.7$), dashed-bold line ($d=7$ and
$\alpha =0.3$).} \label{Fig10}
\end{figure}

\begin{figure}[tbp]
$%
\begin{array}{ccc}
\epsfxsize=5cm \epsffile{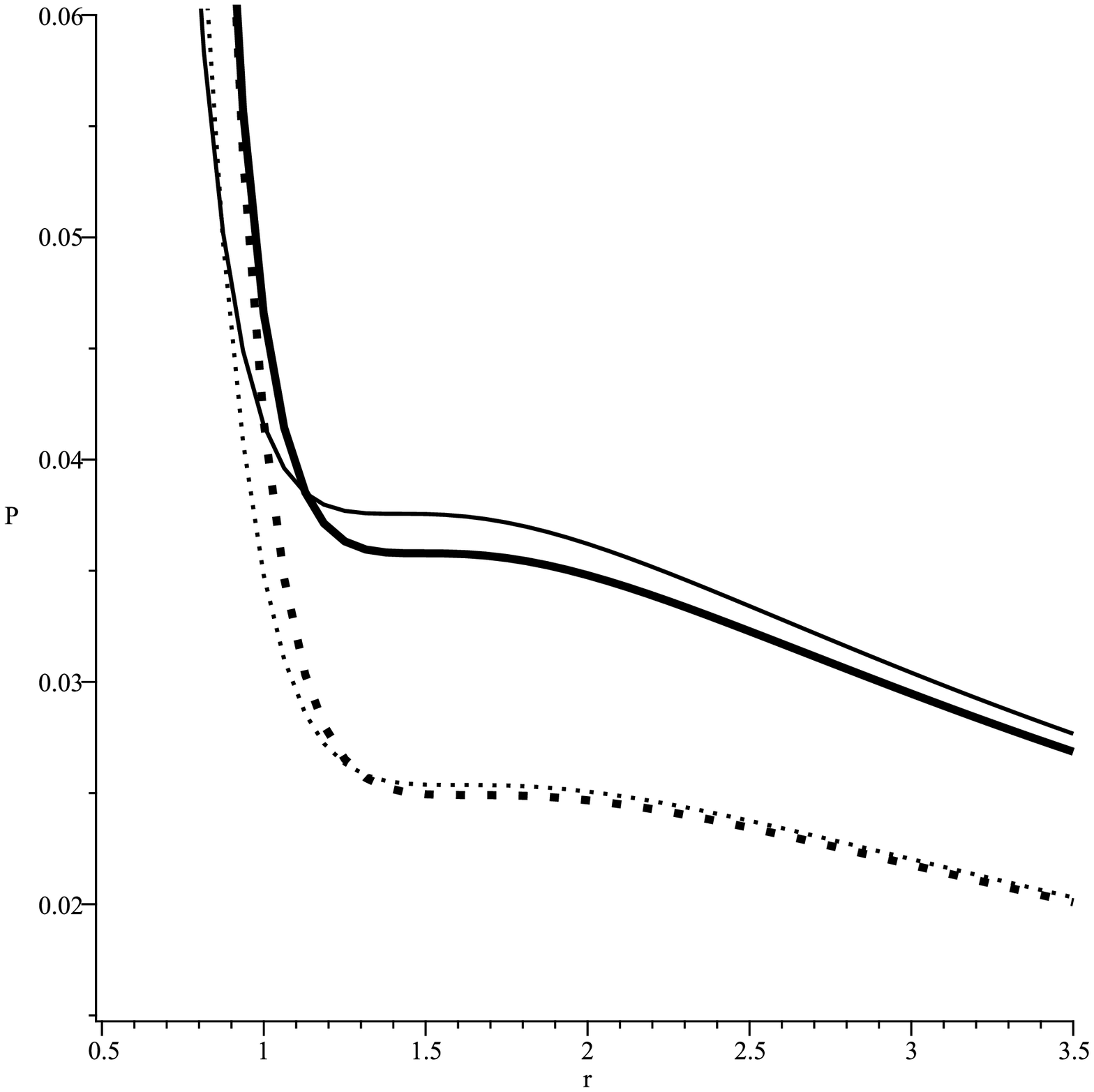} & %
\epsfxsize=5cm \epsffile{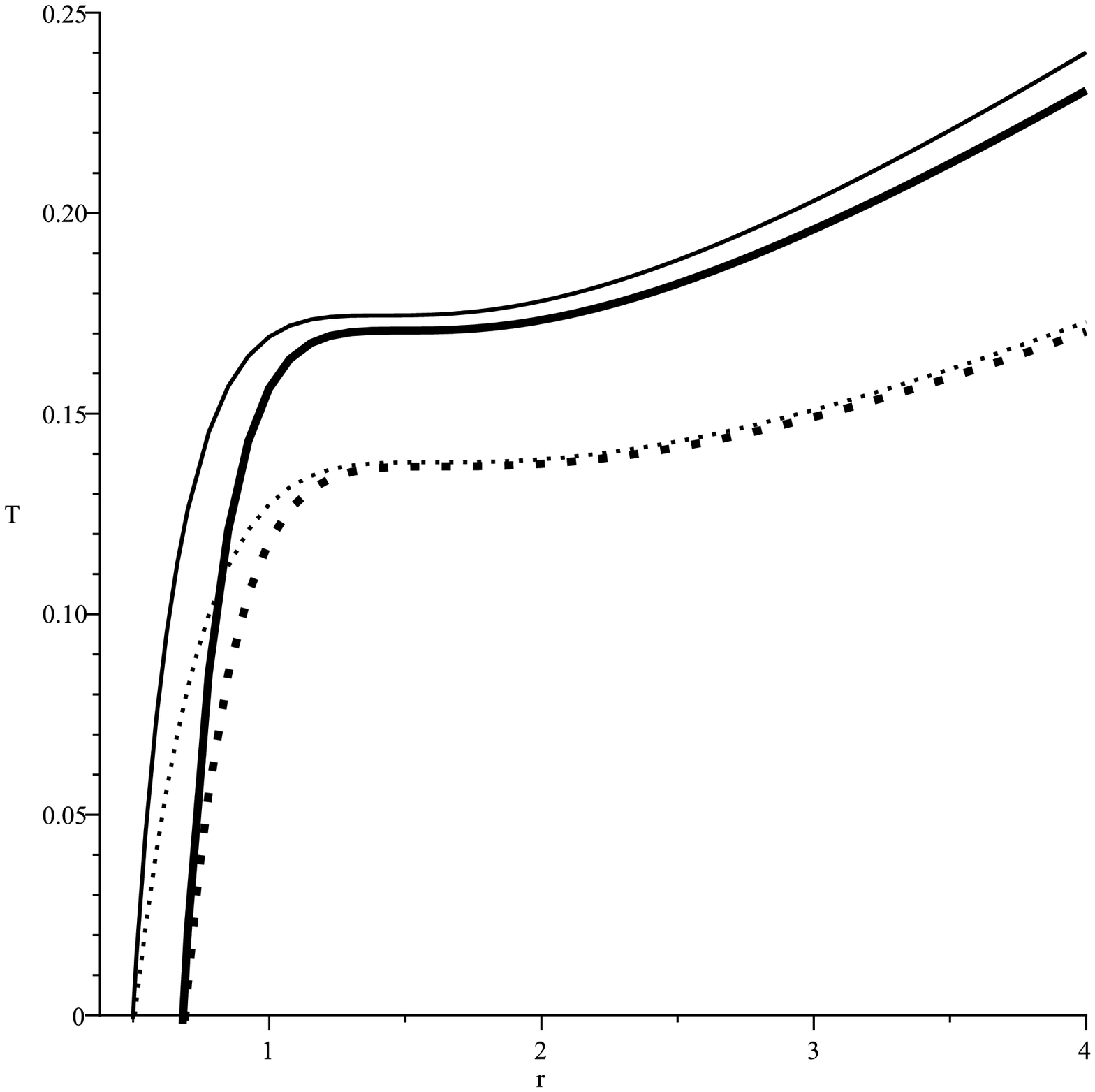} & %
\epsfxsize=5cm \epsffile{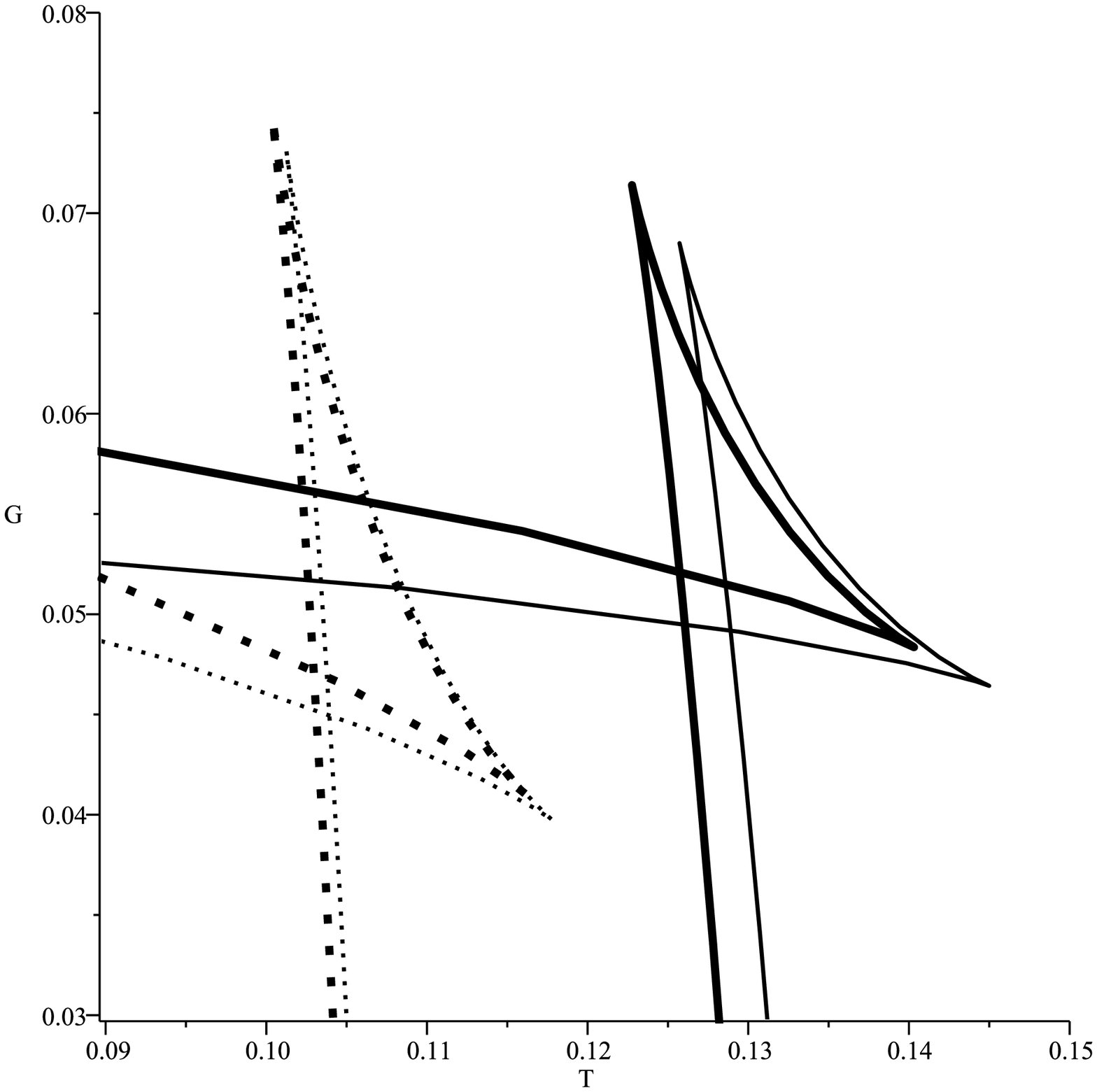}%
\end{array}
$%
\caption{$P-v$ diagram for $T=T_{c}$ (left), $T-v$ diagram for
$P=P_{c}$ (middle) and $G-T$ diagram for $P=0.5P_{c}$ (right) with
$q=1$ and $d=5$. \newline continues line (EN gravity with
$\beta=0.5$), continues-bold line (EN gravity $\beta=1.5$), dotted
line (GB gravity with $\beta=0.5$), dotted-bold line (GB gravity
$\beta=1.5$)} \label{Fig11}
\end{figure}

\begin{figure}[tbp]
$%
\begin{array}{ccc}
\epsfxsize=5cm \epsffile{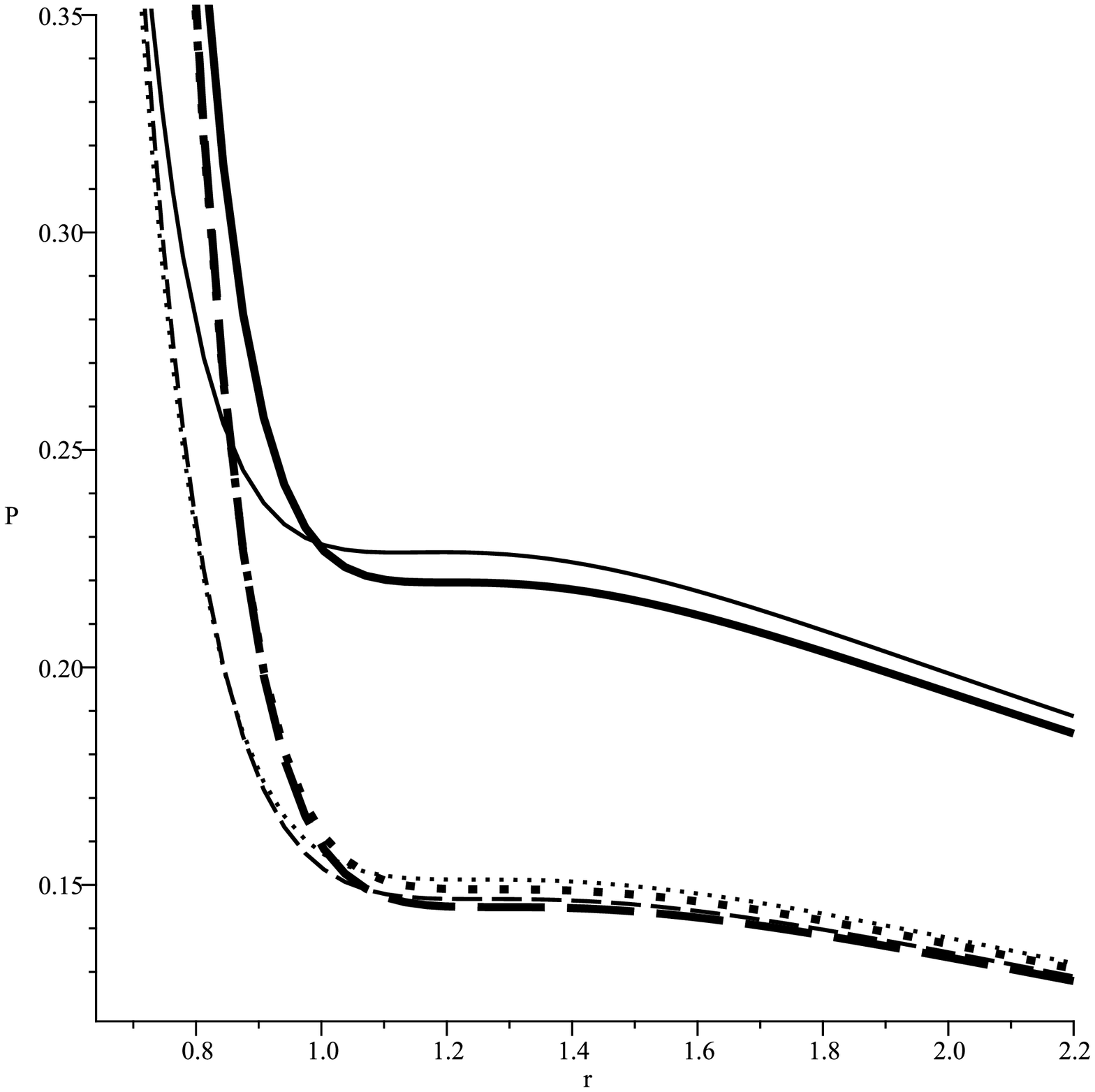} & %
\epsfxsize=5cm \epsffile{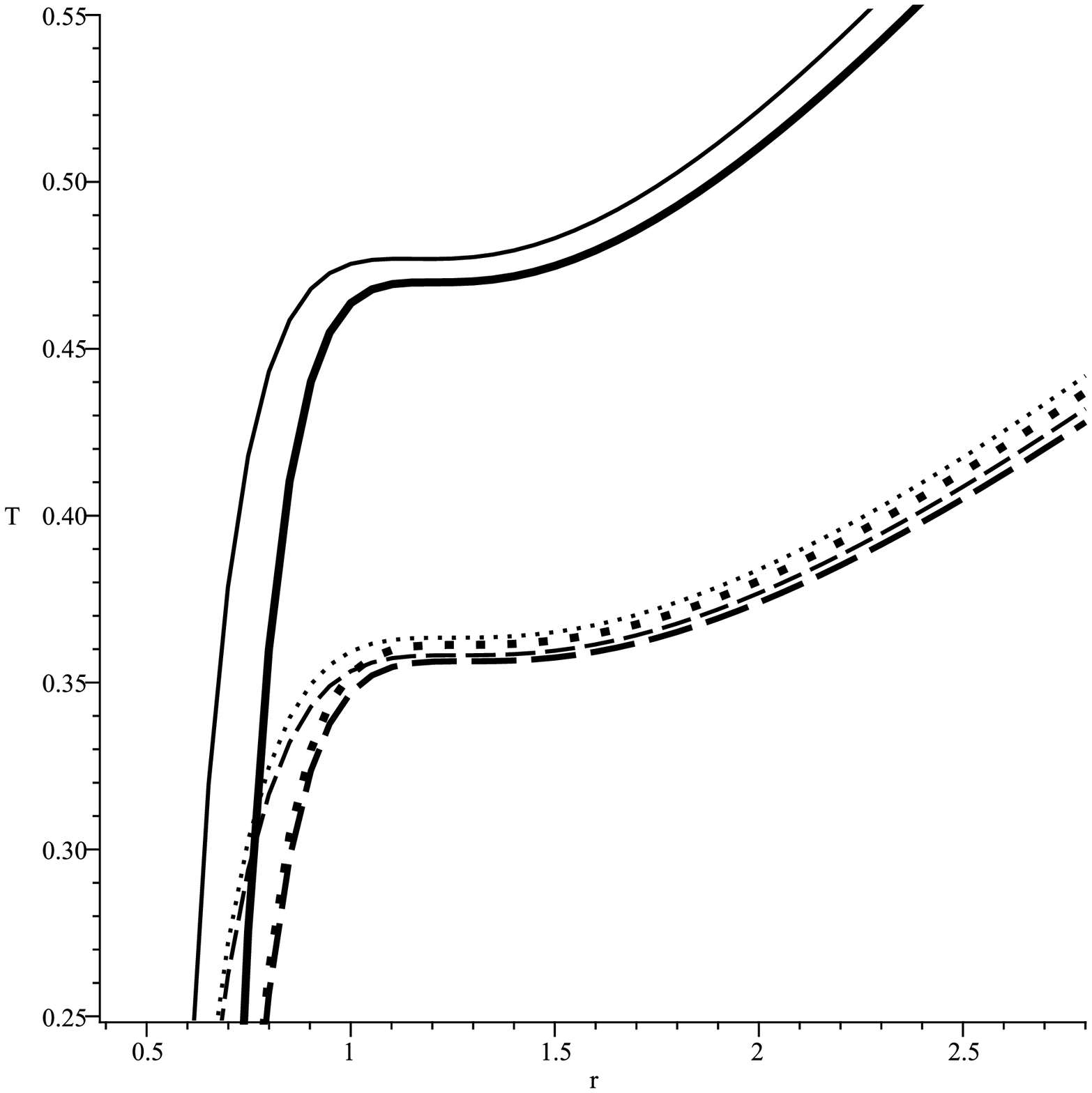} & %
\epsfxsize=5cm \epsffile{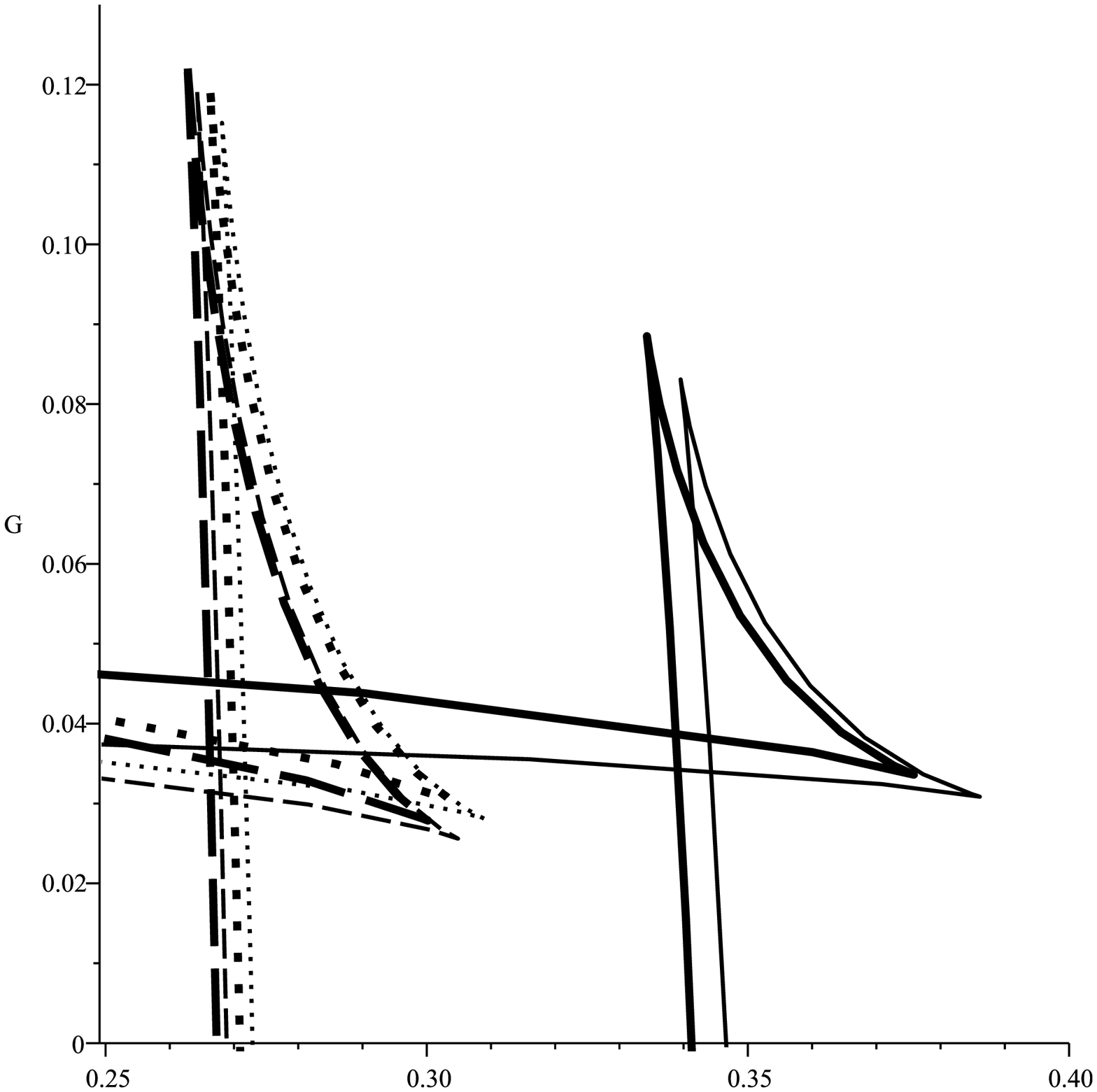}%
\end{array}
$%
\caption{$P-v$ diagram for $T=T_{c}$ (left), $T-v$ diagram for
$P=P_{c}$ (middle) and $G-T$ diagram for $P=0.5P_{c}$ (right) with
$q=1$ and $d=7$. \newline continues line (EN gravity with $\beta
=0.5$), continuous-bold line (EN gravity with $\beta =1.5$),
dotted line (GB gravity with $\beta =0.5$), dotted-bold line (GB
gravity with $\beta =1.5$), dashed line (TOL gravity with $\beta
=0.5$) and dashed-bold line (TOL gravity with $\beta =1.5$).}
\label{Fig12}
\end{figure}

In addition, the energy gap between two phases, critical
temperature and Gibbs free energy are increasing functions of
dimensions (Fig. \ref{Fig10}). In other words, for higher
dimensions, system needs to have more energy for having phase
transition. It is worthwhile to mention that subcritical isobars
(also critical region) are increasing functions of dimensions (see
Figs. \ref{Fig10} middle panels). Whereas critical volume is a
decreasing function of dimensions. Finally, the ratio
$\frac{P_{c}v_{c}}{T_{c}}$ is a decreasing (an increasing)
function of $\alpha$ ($\beta$).



\section{Conclusions}

In this paper, we have considered both GB and TOL gravities in
presence of two classes of BI type NED and studied their phase
diagrams. We have considered cosmological constant as pressure and
its conjugating quantity as thermodynamical volume. The obtained
volume for these cases was consistent with topological structure
of black holes and what was obtained previously \cite{HPE}. It was
shown that although both higher orders of Lovelock gravity and NED
modify thermodynamical quantities but the volume of the black
holes in these cases is independent of these two modifications and
only depends on the topology of the solutions. By employing
numerical method, we have calculated critical thermodynamical
values for different cases and studied the effects of
gravitational and nonlinear electromagnetic field parameters on
these critical values. It was shown that black holes under
consideration have similar behavior as Van der Waals liquid-gas
system.

We found that critical temperature and pressure were decreasing
functions of orders and/or coefficients of Lovelock gravity and
critical volume and energy gap were increasing functions of them.
In other words, black holes with higher orders of Lovelock gravity
go under phase transition and acquire stable state with lower
temperature comparing to Einstein case. On the other hand, the
length of the subcritical isobars and region of the phase
transition were increasing functions of the orders and/or
coefficients of Lovelock gravity. Orders of the Lovelock gravity
are denoting different powers of the curvature scalar. From what
we have obtained, one can argue that the critical temperature,
pressure and the region of the small/large black holes are
decreasing functions of the power of the curvature scalar. While,
the critical volume, subcritical isobars and region of the phase
transition are increasing functions of it. Therefore, the power of
the curvature scalar indeed has a crucial role in variation of the
critical values. It is also regarded that considering the higher
orders of Lovelock theory, increases the power of gravity.

It was shown that critical temperature and pressure were
decreasing functions of $\beta$ whereas the critical volume is an
increasing function of it. In comparison between BI type and
Maxwell electrodynamics, it was found that the lowest critical
pressure and temperature and the largest critical volume belong to
linear (Maxwell) theory.

The gravitational and electromagnetic fields have opposite effects
on critical pressure and also phase transition. According to these
results, one can say that phase transition is fundamentally
related to both gravity and electrodynamics and their powers. As
the power of gravitational field (electromagnetic field) increases
(decreases) the critical temperature decreases (increases).
Interaction between the gravitational and electrodynamic sectors
of charged black hole may be found through the metric function. In
addition, investigations of the phase diagrams confirm that they
are weakening each others effects.

As for dimensionality, we found that as it increases, the energy
gap, critical temperature and critical pressure increase. The
universal ratio of $\frac{P_{c}v_{c}}{T_{c}}$ was a decreasing
function of $\beta$ and increasing function of dimensions and
Lovelock parameter.

For higher orders of Lovelock gravity we have higher value of
entropy ($\alpha >0$) which indicates that the thermodynamical
system that the gravity describe contains higher value of
disorder. In addition, considering higher orders of Lovelock
gravity will cause the black holes to have higher degree of
complexity in their geometrical structure. If one considers the
complexity of the geometrical structure as a disorder measurement
of the system, it is logical to expect to see higher value of
entropy for higher order of Lovelock gravity. On the other hand,
higher value of entropy means that our systems will have phase
transition in lower critical temperature. That is the result that
we have derived through our numerical calculations.

It was shown that \cite{CaiGB} GB gravity in the presence of
Maxwell field has no phase transition for arbitrary electric
charge in higher than $5$-dimensions. While we found that
adjusting the nonlinearity parameter, $\beta$, there is phase
transition for various values of $q$ and $d\geq 5$. In other
words, the nonlinearity parameter of electrodynamics has modified
both electrodynamics and thermodynamical behavior of a black hole
system.

Due to the opposite effects of gravitational power in Lovelock
gravity and the power of nonlinearity in electrodynamics, it is
constructive to find the dominant effect for various domains of
thermodynamical systems. Also, one can study whether these two
different fields for certain values of $\alpha$ and $\beta$,
cancel each other effects or not.

Considering the effects of Hawking radiation, we expect to see
different behavior for higher orders of Lovelock gravity. This
indicates that in order to investigate black holes evaporation and
phase transition of black holes, one could take both effects
simultaneously. Another interesting issue is studying the
connection between complexity of the spacetime (topological
structure of black hole) and entropy of the system and the
interpretation of entropy as geometrical property. We leave these
problems for the future work.

\begin{acknowledgements}
We thank the Shiraz University Research Council.
This work has been supported financially by the Research Institute
for Astronomy and Astrophysics of Maragha, Iran.
\end{acknowledgements}

\end{document}